\documentclass[12pt,reprint,aps,groupedaddress,showpacs,nofootinbib,prd]{revtex4-1}
\usepackage{amsmath,amssymb,graphicx,amsfonts}
\usepackage{mathrsfs}
\usepackage{xcolor}
\usepackage[shortlabels]{enumitem}
\usepackage[utf8]{inputenc}

\newcommand{\orr}{\omega_{r}}
\newcommand{\ot}{\omega_{\theta}}
\newcommand{\veff}{V_{\text{eff}}}
\newcommand{\nut}{\nu_{\theta}}

\newcommand{\TT}{\text{TT}}

\usepackage{hyperref}
\usepackage{setspace}

\begin{document}

\title{Testing spacetime symmetry through gravitational waves from extreme-mass-ratio inspirals}

\author{Kyriakos Destounis}
\email{kyriakos.destounis@uni-tuebingen.de}
\affiliation{Theoretical Astrophysics, IAAT, University of Tübingen, 
72076 Tübingen, Germany}

\author{Arthur G. Suvorov}
\affiliation{Theoretical Astrophysics, IAAT, University of Tübingen,  
72076 Tübingen, Germany}

\author{Kostas D. Kokkotas}
\affiliation{Theoretical Astrophysics, IAAT, University of Tübingen,
72076 Tübingen, Germany}

\date{\today}
 
\begin{abstract}
\noindent{One of the primary aims of upcoming space-borne gravitational wave detectors is to measure radiation in the mHz range from extreme-mass-ratio inspirals. Such a detection would place strong constraints on hypothetical departures from a Kerr description for astrophysically stable black holes. The Kerr geometry, which is unique in general relativity, admits a higher-order symmetry in the form of a Carter constant, which implies that the equations of motion describing test particle motion in a Kerr background are Liouville-integrable. In this article, we investigate whether the Carter symmetry itself is discernible from a generic deformation of the Kerr metric in the gravitational waveforms for such inspirals. We build on previous studies by constructing a new metric which respects current observational constraints, describes a black hole, and contains two non-Kerr parameters, one of which controls the presence or absence of the Carter symmetry, thereby controlling the existence of chaotic orbits, and another which serves as a generic deformation parameter. We find that these two parameters introduce fundamentally distinct features into the orbital dynamics, and evince themselves in the gravitational waveforms through a significant dephasing. Although only explored in the quadrupole approximation, this, together with a Fisher metric analysis, suggests that gravitational wave data analysis may be able to test, in addition to the governing theory of gravity, the underlying symmetries of spacetime.}
\end{abstract}

\maketitle

\section{Introduction}

Many binary merger events involving black holes have been detected by the Laser Interferometer Gravitational-Wave Observatory (LIGO) and Virgo \cite{bhcat}. While a wealth of physical knowledge can and has been extracted from these experiments \cite{ligotests1,ligotests2}, ground-based interferometers are limited by a seismic noise cutoff ($\sim 10$ Hz for Advanced LIGO \cite{ligonoise}), which, due to the fact that the gravitational wave (GW) frequencies emitted by a binary source scale inversely with the constituent masses, prevents them from measuring signals from objects above masses of $\lesssim 100 M_{\odot}$. Space-borne detectors, such as the Laser Interferometer Space Antenna (LISA) and Taiji \cite{lisafreq,taiji}, have the advantage that they are not limited by this noise, and thus can observe higher mass systems in the frequency band $10^{-4} \lesssim f \lesssim 10^{-1} \text{ Hz}$ \cite{Barausse}. One of the main targets for space-based detectors are extreme-mass-ratio inspirals (EMRIs), which are systems involving a super-massive black hole and a plunging, stellar-mass companion \cite{emri}. Incidentally, a major practical advantage in modelling these systems is that the companion can be treated, to high accuracy, as a point particle traversing the gravitational field generated by the super-massive object. GWs from EMRIs are therefore expected to contain detailed information about black-hole structure \cite{barack04,barack07,berti19}. 

In general relativity (GR), astrophysically stable black holes in vacuum, super-massive or otherwise, must be Kerr \cite{Heusler,bhuniq1}. The Kerr metric, being stationary and axisymmetric, admits two Killing vectors which respectively imply that the energies and angular momenta of relativistic particles moving within the spacetime are conserved. Together with the particle Hamiltonian, we then have three constants of motion, though these alone are insufficient to ensure that the equations of motion are (Liouville-)integrable. In this context, integrability implies the absence of chaotic orbital phenomena (e.g. \cite{cont02}). The Kerr spacetime, however, admits a rank-two Killing tensor, which provides a fourth constant of motion in the form of the Carter constant \cite{carter68}, and thus orbits necessarily display regular behaviour \cite{wilkins72}. If, however, GR provides an inexact description of the geometry surrounding compact objects (as anticipated by, e.g., perturbative non-renormalizability \cite{thooft}), or the hole's geometry is significantly warped due the presence of an accretion disk \cite{rez07,rez08,sem13}, it may be that the Kerr metric only approximately describes super-massive black holes (though cf. Refs. \cite{psaltis08,gurl15}). Depending on how potential departures from a Kerr description manifest, the integrability property may be broken. 

Due to the huge mass disparity, EMRIs may survive for up to several years before the plunge. It is therefore possible to model certain parts of the inspiral as bound orbits, which are characterized by two libration-like frequencies, one of which ($\orr$) is radial and describes the rate of transition from the periapsis to the apoapsis of the orbit, and the other ($\ot$) which describes longitudinal oscillations around the equatorial plane \cite{brink13}. When a non-integrable perturbation is introduced into the Hamiltonian, only a finite number of the periodic orbits that had a rational ratio of the above frequencies (`resonant' orbits) survive. Surrounding each surviving periodic orbit, a small `island' of stability appears in the phase space; the orbits in these islands share the same rational ratio of frequencies with the central periodic one without necessarily being periodic \cite{moser62,arnold63,arnold89}. When an inspiraling orbit crosses an island, the ratio $\orr/\ot$ remains constant, whereas it otherwise behaves monotonically as a function of radius, and so the orbital dynamics display transient plateau features. The question then becomes whether or not such a plateau feature imprints a discernible signature onto the GW signal \cite{apo07,luk10,cont11,luk14,berry16}. In any case, an identification of such a feature in upcoming LISA data may hint at physics describing a fundamentally non-Kerr spacetime (though cf. Refs.  \cite{kiu04,spin1,spin2,zil20}). Moreover, Ryan \cite{ryan95,ryan97} has shown that the spacetime multipole moments are retrievable from the Keplerian frequency $\omega_{\phi}$ together with $\ot$ and $\orr$, so that EMRI analysis can be used to test no-hair conjectures experimentally, independently of GR's validity \cite{babak05,Barack:2006pq,Gair:2011ym,suvmel16,Moore:2017lxy,Chua:2018yng}. 

However, the extent to which a plateau feature, which signals chaos, may be distinguishable from more general deformed-Kerr traits is not obvious \cite{cont11,berry16}. For instance, one could imagine that the EMRI waveform associated to a deformed-Kerr object, i.e. one which still respects a generalized Carter symmetry \cite{carter68}, might mimic that of a non-Kerr object sufficiently well to prevent an identification of the underlying spacetime symmetries. It is the aim of this paper, by building on previous studies, to investigate some aspects of this point, that is to see whether GWs from EMRIs can test the fundamental symmetry properties of astrophysical black holes. Although we do not solve a realistic, non-Einstein plunging problem, we investigate GWs associated with bound orbits in a new spacetime to provide some first steps. Specifically, some elementary data analysis relevant for LISA is carried out using Fourier and Hilbert transforms together with a Fisher metric analysis, to characterize the imprint on the resulting waveforms. Recently, Johannsen \cite{joh13} developed a metric which includes generic deviations from the Kerr spacetime while preserving the Carter constant (see also Refs. \cite{brink10,papkok18,car20}) and otherwise remaining regular. Here, we adopt Johannsen's metric \cite{joh13} though introduce an extra parameter which explicitly breaks the Carter symmetry but maintains the other desirable properties. Massive particle orbits and the associated gravitational multipole moments in this spacetime are computed to study how EMRI data analysis can probe spacetime symmetries.

This paper is organized as follows. In Sec. II we introduce the geodesic equations for a general stationary, axisymmetric spacetime, and describe features that may manifest in orbital dynamics if the system is not integrable. In Sec. III, a  particular metric is introduced which contains two free parameters, one of which controls the presence (or absence) of the Carter symmetry, and another which deforms the black hole from the Kerr geometry. Sec. IV then solves the geodesic equations for this spacetime, and compares chaotic features in the orbit with the gravitational waveform associated to an EMRI (Sec. V). Some discussion is presented in Sec. VI.

\section{Orbital dynamics}

A general stationary, axisymmetric metric can be written in Boyer-Lindquist coordinates $(t,r,\theta,\phi)$ as \cite{pap53}
\begin{equation} \label{eq:genmet}
ds^2 = g_{tt} dt^2 + 2 g_{t \phi} dt d \phi + g_{rr} dr^2 + g_{\theta \theta} d \theta^2 + g_{\phi \phi} d \phi^2,
\end{equation} 
where each of the metric components are functions of $r$ and $\theta$ only. Geodesic motion is then governed by the Hamiltonian (e.g. \cite{luk10})
\begin{equation} \label{eq:hamiltonian}
H(\boldsymbol{x}, \boldsymbol{p}) = \frac {1} {2 \mu} g_{\alpha \beta}(\boldsymbol{x}) \dot{x}^{\alpha} \dot{x}^{\beta}
\end{equation}
for a particle of mass $\mu$ with momenta $p^{\mu} = \dot{x}^{\mu}$, where an overhead dot denotes differentiation with respect to the proper time $\tau$. From stationarity and axisymmetry, Hamilton's equations immediately give $\dot{p}_{t} = 0 = \dot{p}_{\phi}$, implying that we have two constants of motion, namely the energy $E$ and angular momentum $L_{z}$. Specifically, the $t$- and $\phi$-momenta are given by
\begin{equation}
\label{constants1}
\dot{t} =\frac{E g_{\phi\phi}+L_z g_{t\phi}}{g_{t\phi}^2-g_{tt}g_{\phi\phi}},
\end{equation}
and
\begin{equation} \label{constants2}
\dot{\phi} =-\frac{E g_{t\phi}+L_z g_{tt}}{g_{t\phi}^2-g_{tt}g_{\phi\phi}},
\end{equation}
and are completely determined by $E$, $L_z$, and the metric components. The remaining equations of motion read
\begin{align}
2\,g_{rr}\,\ddot{r}+2\,\dot{r}\,\dot{\theta}\,\partial_\theta g_{rr}+\dot{r}^2\,\partial_r g_{rr}-\dot{t}^2\,\partial_r g_{tt}-2\,\dot{t}\,\dot{\phi}\,\partial_r g_{t\phi}\nonumber\\
\label{eqr}
-\dot{\theta}^2\,\partial_r g_{\theta\theta}-\dot{\phi}^2\,\partial_r g_{\phi\phi}=0,\\
2\,g_{\theta\theta}\,\ddot{\theta}+2\,\dot{r}\,\dot{\theta}\,\partial_r g_{\theta\theta}-\dot{r}^2\,\partial_\theta g_{rr}-\dot{t}^2\,\partial_\theta g_{tt}-2\,\dot{t}\,\dot{\phi}\,\partial_\theta g_{t\phi}\nonumber\\
\label{eqtheta}
+\dot{\theta}^2\,\partial_\theta g_{\theta\theta}-\dot{\phi}^2\,\partial_\theta g_{\phi\phi}=0.
\end{align}
The constants of motion \eqref{constants1} and \eqref{constants2}, together with the conservation of the particle's rest mass $H=-\mu$, lead to the constraint equation \cite{zelenka17}
\begin{equation} \label{eq:equationsofmotion}
\dot{r}^2 + \frac {g_{\theta \theta}} {g_{rr}} \dot{\theta}^2 + \veff = 0,
\end{equation}
where 
\begin{equation}
\label{veff}
\veff\equiv \frac{1}{g_{rr}} \left( 1 + \frac {g_{\phi \phi} E^2 + g_{tt} L_{z}^2 + 2 g_{t \phi} E L_{z}} {g_{tt} g_{\phi \phi} - g_{t \phi}^2} \right)
\end{equation}
is the \emph{effective potential}. The curve defined by the vanishing of $\veff$ is called the \emph{curve of zero velocity} (CZV), since $\dot{r} = \dot{\theta} = 0$ there, implying that whenever a geodesic orbit reaches this surface the velocity components in the $(r,\theta)$-plane vanish \cite{cont02}.

\subsection{Signatures of non-integrability}

As evidenced by expression \eqref{eq:equationsofmotion}, the dynamics of geodesic motion for the metric \eqref{eq:genmet} can be reduced to motion on a two-dimensional surface, where orbits oscillate in the two available degrees of freedom with characteristic frequencies $\orr$ and $\ot$. The ratio of these two quantities, $\nut = \orr / \ot$, is called the \emph{rotation number}, and carries pertinent information about the properties of the trajectory. Specifically, when the particle path intersects a particular surface of section within the phase space $(r, \theta, \dot{r}, \dot{\theta})$ [e.g. the two-dimensional sub-space spanned by $(r,\dot{r})$ for $\theta = \pi/2$, $\dot{\theta} = 0$], one can track the angles $\vartheta$ formed between subsequent intersection points relative to some fixed center $\boldsymbol{u}_{0}$. Denoting the $n^{\text{th}}$ crossing of the orbit through a surface of section by $\boldsymbol{w}_{n}$, the angles $\vartheta_{n+1} = \arg\left(\boldsymbol{R}_{n+1},\boldsymbol{R}_{n}\right)$ for position vectors $\boldsymbol{R}_{n} = \boldsymbol{w}_{n} - \boldsymbol{u}_{0}$ track the points of intersection and define a sequence of rotation numbers, 
\begin{equation} \label{eq:nun}
\nu_{\theta,N} = \frac {1} {2 \pi N} \sum^{N}_{i=1} \vartheta_{i}.
\end{equation}
In the limit $N \rightarrow \infty$, the sequence \eqref{eq:nun} converges to $\nut$. In general, regular behavior is characterized by monotonicity in high-$N$ rotation numbers with respect to an increase in one of the phase space parameters while others stay fixed; this essentially states that, as one smoothly approaches the boundary of the CZV, the orbits do not behave erratically in an integrable system (see Figure 2 in Ref. \cite{lg12} for a schematic representation of these quantities).

The Kerr spacetime, in addition to admitting the invariants $\mu$, $E$, and $L_{z}$ discussed above, has a `hidden' symmetry in the form of a rank-2 Killing tensor, giving rise to the Carter constant \cite{Heusler,carter68}, which is quadratic in the momenta $\boldsymbol{p}$. This implies that the equations of motion \eqref{constants1}-\eqref{eqtheta} are integrable, and thus that the high-$N$ rotation numbers on any given surface of section for the Hamiltonian \eqref{eq:hamiltonian} behave monotonically with increasing radius. Moreover, the orbit is periodic for rational rotation numbers \cite{arnold89}. When $\nut$ is irrational, the motion is instead quasi-periodic and densely covers an invariant torus within the phase space $(r, \theta, \dot{r}, \dot{\theta})$. 

In general, the Kolmogorov-Arnold-Moser (KAM) theorem \cite{moser62,arnold63} states that, when introducing a non-integrable perturbation $H_{1}$ into the Kerr or any other regular Hamiltonian $H_{0}$, i.e. for $H = H_{0} + \epsilon H_{1}$, the dynamics will smoothly deviate from the background dynamics provided that the rotation numbers are `sufficiently irrational' so as to satisfy Arnold's criterion \cite{arnold63},
\begin{equation} \label{eq:arnold}
|n \orr - m \ot| > \frac {K(H_{0},\epsilon)} {(n+m)^3},
\end{equation}
where the prefactor $K$ is some complicated function that approaches zero as $\epsilon \rightarrow 0$. However, when $m$ is an integer multiple of $n$ or vice versa, the orbit is resonant and the dynamics are fundamentally altered: the Poincar{\'e}-Birkhoff theorem tells us that exactly half of these orbits become unstable while the other half remain stable \cite{poin12,birk13}. Small islands of stability (`Birkhoff islands') form around certain intersection points $\boldsymbol{w}$ on the surface of section, and exhibit the crucial feature that $\nut$ is constant when an orbit traverses an island, thereby forming a plateau in the graph of $\nut$ \cite{cont02}.

While chaotic orbits are difficult to track (since the onset of chaos need not begin until an arbitrarily large number of cycles have elapsed), the identification of plateaus in the rotation number profile are sufficient to identify an absence of a (generalized) Carter constant as they only appear in non-integrable systems \cite{apo07,luk10,cont11}. 

\section{Deformed and non-Kerr black holes}

In this paper, we are interested in identifying whether gravitational waveforms associated with EMRIs can be used to identify the underlying symmetries of the spacetime. To achieve this, we introduce a new metric which, although not arising as an exact solution within any known theory of gravity (though see Ref. \cite{suv20}), generalizes the Kerr metric by admitting two extra parameters, one of which controls the integrability of the geodesic equations, and the other which deforms the geometry but maintains integrability. 

Although there are a number of such eponymous metrics, the Johannsen metric \cite{joh13} [see equations (51) therein] is a generalization of the Kerr metric designed such that a number of desirable properties are maintained, most notably that the spacetime admits a Carter constant. As such, a stationary and axisymmetric metric which cannot be cast into Johannsen's most general form (see also Refs. \cite{papkok18,car20}) will not possess the Carter symmetry\footnote{We note that this does not automatically preclude integrability; a spacetime might, in principle, admit integrals of motion which are cubic or higher-order in $\boldsymbol{p}$, though this does not happen in our case. Furthermore, there is evidence to suggest this is not possible for stationary and axisymmetric spacetimes \cite{krug15,voll17}.}. There are many ways one can build such a metric, almost all of which will introduce pathological features into the spacetime, e.g. signature changes outside of the horizon(s).

Nevertheless, consider the metric \eqref{eq:genmet} with components
\begin{equation}\label{eq:metric}
\begin{aligned}
&g_{tt} = -\frac{\Sigma[\left(\alpha_{Q}/r\right)M^3  + \Delta-a^2 A(r)^2 \sin^2\theta]}{[(r^2+a^2)-a^2 A(r) \sin^2\theta]^2}, \\
&g_{t\phi} = -\frac{a [(r^2+a^2)A(r)-\Delta] \Sigma \sin^2\theta}{[(r^2+a^2)-a^2 A(r)\sin^2\theta]^2},  \\
&g_{rr} = \frac{\left(\alpha_{Q}/r\right)M^3  + \Sigma}{\Delta} , \\
&g_{\theta \theta} = \Sigma,  \\
&g_{\phi \phi} = \frac{\Sigma \sin^2 \theta \left[(r^2 + a^2)^2 - a^2 \Delta \sin^2 \theta \right]}{[(r^2+a^2) -a^2 A(r) \sin^2 \theta]^2},
\end{aligned}
\end{equation}
for
\begin{equation}
\Sigma = r^2 + a^2 \cos^2 \theta, 
\end{equation}
\begin{equation}
\Delta = r^2 - 2 M r + a^2,
\end{equation}
and
\begin{equation}
A(r) = 1 + \frac {\alpha_{22} M^2} {r^2},
\end{equation}
where $M$ and $a$ represent the mass and spin of the black hole, respectively, while $\alpha_{22}$ and $\alpha_{Q}$ are deformation parameters. 

The parameter $\alpha_{22}$ is one of the Johannsen deformation parameters that couples to the spin (i.e. trivially vanishes when $a=0$), and thus is likely to have the strongest effect for observables pertaining to rapidly rotating objects of interest: gradual accretion is expected to spin-up a black hole to a maximum value of $a_{\text{max}} \approx 0.998 M$ \cite{thorne74}, so that many super-massive objects in active galactic nuclei, relevant for EMRIs detectable by LISA, may be spinning close to this limit (e.g. \cite{berti08}). 

The other (new) parameter $\alpha_{Q}$ controls the Carter symmetry; for any $\alpha_{Q} \neq 0$ the geodesic equations \eqref{constants1}-\eqref{eqtheta} are not integrable for a spinning object. In any case, if we treat both $\alpha$ parameters as small relative to the background terms, the KAM and Poincar{\'e}-Birkhoff theorems discussed in the previous sections apply. Small or otherwise, the metric \eqref{eq:metric} boasts several nice properties expected of astrophysical black holes: (i) The metric possesses an event horizon at the roots of $\Delta = 0$, an ergosphere at the greatest positive root of $g_{tt} = 0$, and is regular everywhere outside these surfaces. (ii) The metric has a Kerr limit when both free parameters vanish. (iii) The metric is asymptotically flat, and (iv) The metric has a Newtonian limit and, by construction, the post-Newtonian Eddington-Robertson-Schiff parameters $(\gamma, \beta)$ exactly match the GR values for any choice of the free parameters \cite{will18}.

Some intuition about the physical interpretation of the parameters $\alpha_{Q}$ and $\alpha_{22}$ can be gained by comparing the spacetime \eqref{eq:metric} with known exact solutions. For example, the leading-order $tt$-component of a particular non-vacuum black hole metric introduced by Bardeen reads $g_{tt} = -1 + 2 M /r - 3 M q^2/r^3 + \mathcal{O} (1/r^4)$ \cite{bar68}. The parameter $q$ appearing in Bardeen's metric can be interpreted as a magnetic charge \cite{bardeenf}, as the metric solves the Einstein equations coupled to a non-linear electromagnetic field sourced by a monopole. In our case, we have from \eqref{eq:metric} that $g_{tt} = -1 + 2 M /r - M^3 \alpha_{Q}/r^3$ in the static limit, and therefore the parameter $\alpha_{Q}$ can be thought of as a gravitational analogue of a Bardeen magnetic monopole. The parameter $\alpha_{22}$ on the other hand couples to the spin at next-to-leading order in $g_{t \phi}$, and may therefore be thought of as modulating the extent of frame-dragging in the spacetime. More formal investigations into parameter interpretations are possible with tools like multipole moment expansions (see Section XI of \cite{thorne80}), though utilizing these schemes is challenging in practice (cf. Ref. \cite{konop16}).

\section{Geodesic orbits and chaos}

In this section, we investigate the orbital dynamics of a bound, test particle of mass $\mu$ for the spacetime metric \eqref{eq:metric}. This is achieved by numerically integrating equations \eqref{eqr} and \eqref{eqtheta} with respect to $\tau$ for some appropriate set of initial conditions. We take the ratio $\mu/M=10^{-6}$ throughout all our evolutions, and also fix the particle's orbital parameters $E=0.95\mu$ and $L_z=3M\mu$ to simulate nearly circular orbits. The momentum constraint \eqref{eq:equationsofmotion} is checked at each time step to ensure numerical accuracy. For all simulations presented herein, we find that they are satisfied to within one part in $\sim 10^{9}$ for the first $\sim 10^{3}$ crossings through the equatorial plane. 

\begin{figure*}[t]
\includegraphics[scale=0.187]{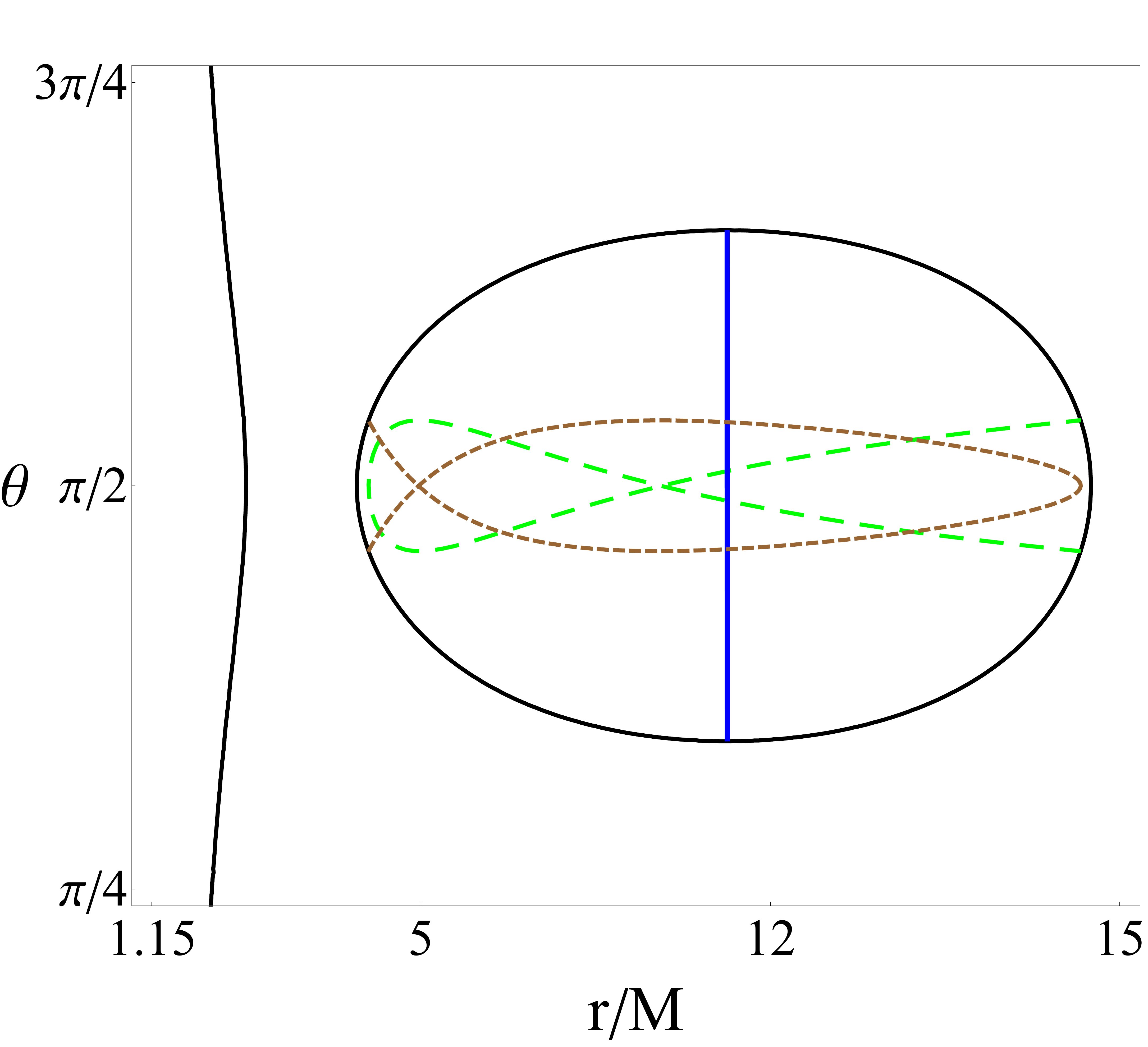}\hskip 6ex
\includegraphics[scale=0.187]{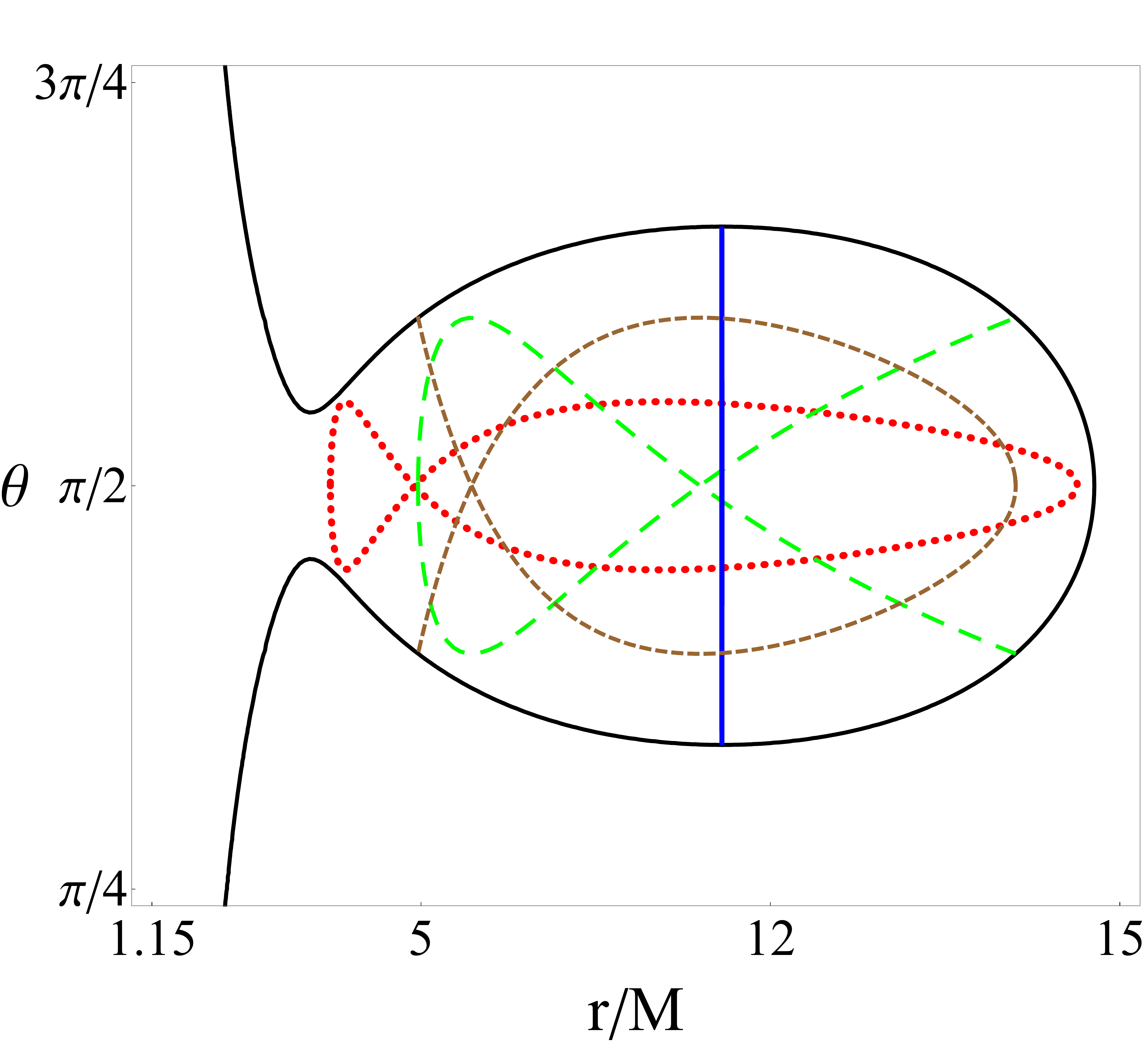}
\caption{Curves of zero velocity (black solid curves) and bound orbits of a test particle with energy $E=0.95\mu$, angular momentum $L_z=3M\mu$, and mass ratio $\mu/M=10^{-6}$ around a non-Kerr black hole with angular momentum $a=0.99M$, for $\alpha_{22}=0$ and $\alpha_{Q}=-1.2$ (left panel), $\alpha_{Q}=-1.8$ (right panel). The red (dotted) curve corresponds to an orbit of $1/2-$resonance and the green (strongly dashed) and brown (lightly dashed) curves correspond to orbits of $2/3-$resonance, while the blue (solid horizontal) lines correspond to orbits which pierce the equatorial plane exactly once. The event horizon is located at the leftmost edge of the $r-$axis.}
\label{CZV}
\end{figure*}

\subsection{Non-Kerr spacetime}

When the Carter symmetry is broken, the geodesic equations for the spacetime described by \eqref{eq:metric} are non-integrable. As such, the CZV, and hence surfaces of section and rotation numbers, deviate from that of Kerr when $\alpha_Q \neq 0$ in some fundamental respects. For $\alpha_Q>0$, the CZV volume decreases with increasing $\alpha_{Q}$ (for fixed $E$ and $L_{z}$), and the test particle has access to only a subset of the bound orbits available to a particle orbiting a Kerr object. On the other hand, when $\alpha_Q<0$ the CZV volume increases, and the test particle has access to a larger region of bound orbits. In this section, we focus on the latter case, since setting $\alpha_Q < 0$ unlocks orbits with small periodicities. This works in our favor since transient phenomena in the evolution of resonant periodic orbits with small rotation numbers are more prominent and easier to capture numerically \cite{luk10}.

In Fig. \ref{CZV} some illustrative examples of the CZV are shown. For $\alpha_Q=-1.2$ (left panel), the CZV forms a closed region of bound motion, as for Kerr with $a \leq M$, though is larger in volume. A second region of plunging orbits is also formed close to the event horizon. For $\alpha_Q=-1.8$ (right panel), the CZV volume is greater than its $\alpha_{Q} = -1.2$ counterpart, and the two regions are connected by a `throat', similar to what is seen for `bumpy' spacetimes \cite{gair08}. The particulars of the initial conditions determine whether an orbit will plunge into the black hole or remain bound around it. Our numerics indicate that for an orbit to be bound, the initial position of the test particle, in the equatorial plane, must not lie at or beyond the throat opening or very close to the maximum radius of the CZV. For both cases, periodic orbits with resonances (i.e. rotation numbers) $1/2$ (red) and $2/3$ (green, brown) are also shown. Such orbits exhibit turning points within the interior of the CZV, in contrast to non-resonant orbits (not shown), all of which have turning points that lie on the boundary of the CZV. A special case of an ordered, non-resonant orbit that pierces the equatorial plane exactly once, thus defining the fixed center $\boldsymbol{u}_0$, is shown in blue. 

\begin{figure*}[t]
\includegraphics[scale=0.18]{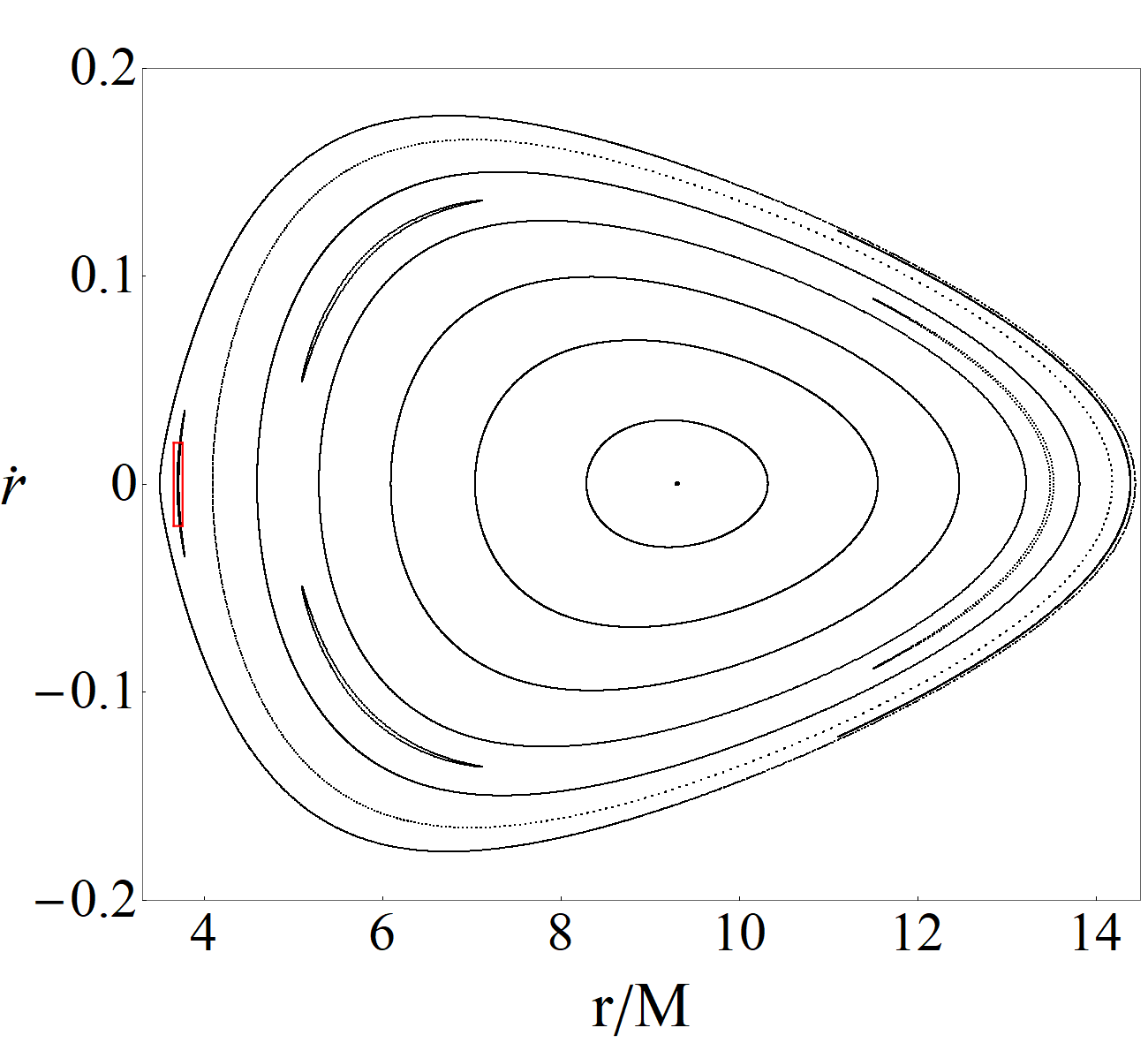}\hskip 6ex
\includegraphics[scale=0.183]{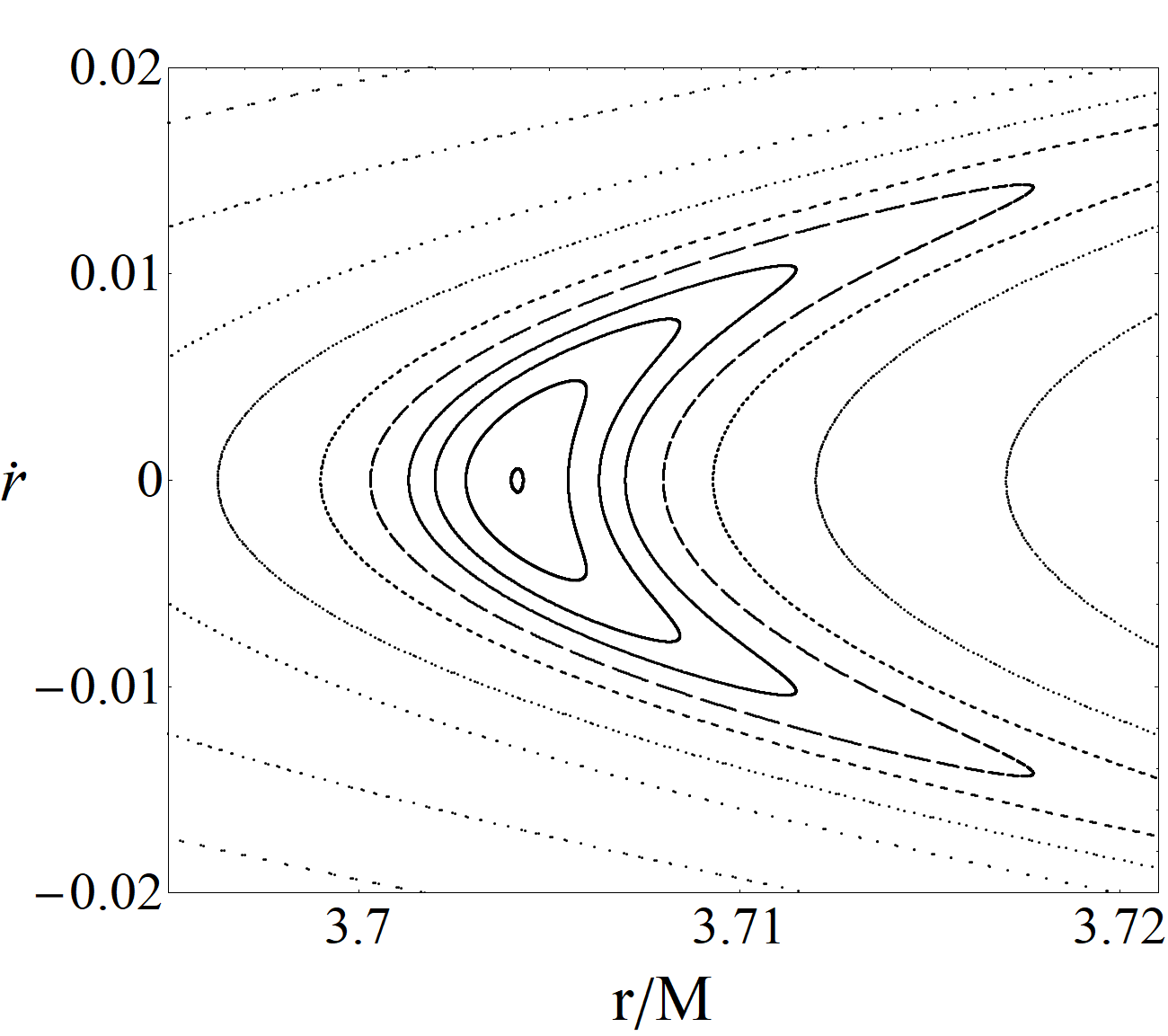}
\caption{Left: The equatorial $(\theta=\pi/2)$ surface of section for the bound orbits of a test particle with energy $E=0.95\mu$, angular momentum $L_z=3M\mu$ and mass ratio $\mu/M=10^{-6}$ around a non-Kerr object with angular momentum $a=0.99M$, for $\alpha_Q=-1.8$ and $\alpha_{22}=0$. Right: The boxed (red) region from the previous surface of section, where one of the Birkhoff chains of islands is more visible. The Birkhoff chain shown is associated with stable periodic orbits of $1/2-$resonance.}
\label{maps}
\end{figure*}

A more practical way of distinguishing between ordered, periodic, and chaotic orbits is by inspecting the Poincar\'e maps, i.e. surfaces of section which trace points where orbits intersect some particular phase subspace. The successive intersections of ordered orbits lie along curves which encircle the fixed center $\boldsymbol{u}_0$, while the intersections of resonant orbits form chains of islands of stability which do not surround this point. The successive iterates of chaotic orbits are scattered irregularly. Fig. \ref{maps} shows such a Poincar\'e map in the equatorial plane for the particular case $\alpha_{Q} = -1.8$, where the dot in the center of the left panel shows the fixed central point $\boldsymbol{u}_{0}$. A zoom-in of the Birkhoff island of resonance $1/2$ is shown in the right-hand panel, which reveals the nesting property common to islands surrounding resonant geodesics. Points exist between the successive islands of stability where unstable orbits of resonance $1/2$ and $2/3$ emanate. Around these unstable points, the orbits are chaotic and their intersections on the Poincar\'e map are irregularly scattered. Such chaotic orbits form an extremely thin layer, which surrounds the Birkhoff islands and are not visible in Fig. \ref{maps}. 

\begin{figure*}[t]
\includegraphics[scale=0.18]{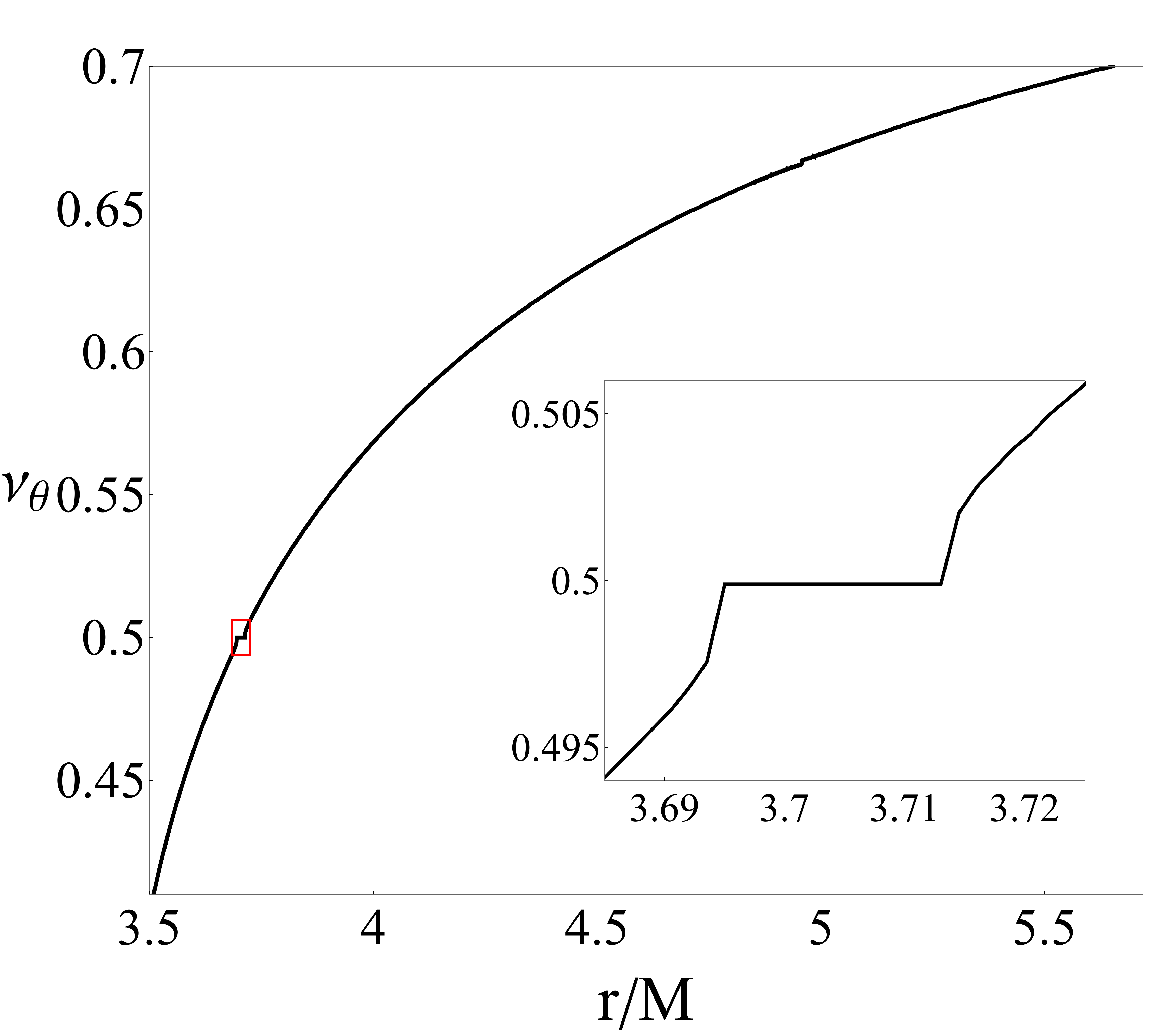}\hskip 5ex
\includegraphics[scale=0.184]{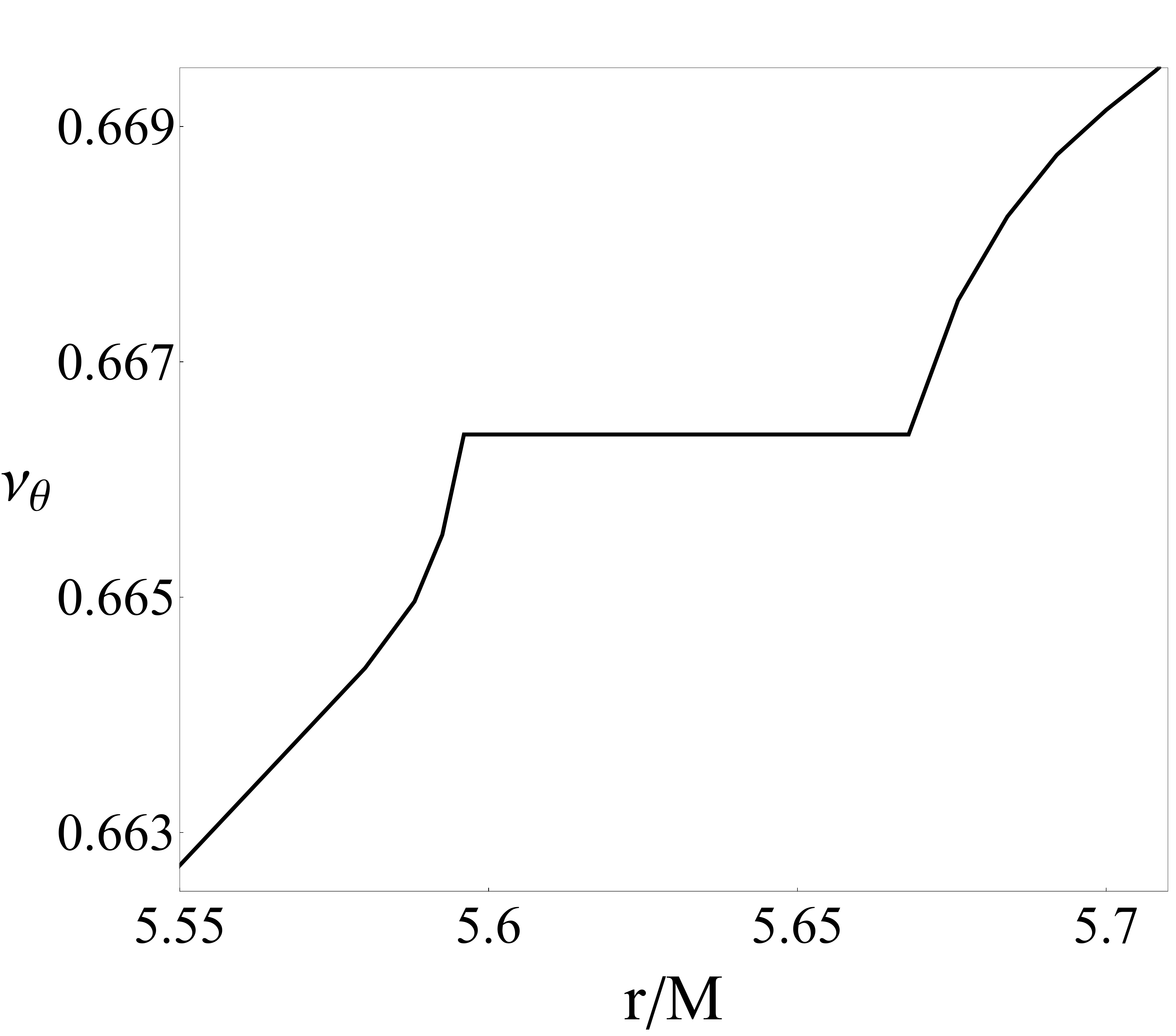}
\caption{Left: The rotation curve as a function of radius along the line $\dot{r}=0$ for the surface of section presented in Fig. \ref{maps}. The plateau presented in the inlay shows the boxed (red) region around the $1/2-$resonance. Right: The rotation curve for the surface of section presented in Fig. \ref{maps} along the line $\dot{r}=0.1$. The plateau is associated with the stable periodic orbits of $2/3-$resonance.}
\label{rotation_aQ}
\end{figure*}

A systematic way of discovering islands of stability is by calculating the rotation numbers, which we show in Fig. \ref{rotation_aQ} as a function of radius for some initial momenta $\dot{r}$, along successive orbits at various distances from the fixed center $\boldsymbol{u}_0$ using the method described in Sec. 2 for some large $N$. In general, for an integrable system such as Kerr geodesics, the rotation curve is strictly monotonic and smooth (cf. the solid curve in Fig. \ref{rotation_a22} below). For $\alpha_Q \neq 0$ the rotation curve exhibits plateaus, designating the crossing into a Birkhoff chain of stable resonant orbits, and inflection points, designating the crossing into an unstable point where chaotic orbits emanate and surround the associated islands of stability. For $\dot{r} =0$ (left panel of Fig. \ref{rotation_aQ}) we can clearly distinguish a plateau associated with the $1/2-$resonant orbit, where $\nu_{\theta} = 1/2$, and an inflection point associated with the $2/3-$resonant orbit, where $\nu_{\theta}=2/3$. The Birkhoff islands of $2/3-$resonance, which lie radially before the fixed center, do not cross the $\dot{r}=0$ axis and therefore cannot be accessed unless $\dot{r}\neq 0$. In the right-hand panel of Fig. \ref{rotation_aQ}, a plateau associated with the $2/3-$resonant orbit can, indeed, be seen along the $\dot{r}=0.1$ line. It is worth noting that, however, the $2/3-$resonant orbit occurring at large ($r/M \sim 13$) radii (that lie beyond the fixed center) can still be observed in the $\nu_{\theta}$ curve for $\dot{r}=0$, though is not shown (see Fig. \ref{maps}).

Sizes and locations of resonant islands of stability depend on both the physical parameters of the metric \eqref{eq:metric} and the parameters of the orbit itself. Our analysis finds that the most prominent islands (the ones with the largest width) are those with the smallest multiplicities (i.e. $2/3-$resonance), as expected.  A similar conclusion was found in Ref. \cite{luk10} for the Manko-Novikov spacetime; see also Ref. \cite{zil20}, where prominence is shown to be related to the proximity of the island to the neighboring chaotic sea.

\subsection{Deformed Kerr spacetime}

\begin{figure}[h]
\includegraphics[scale=0.18]{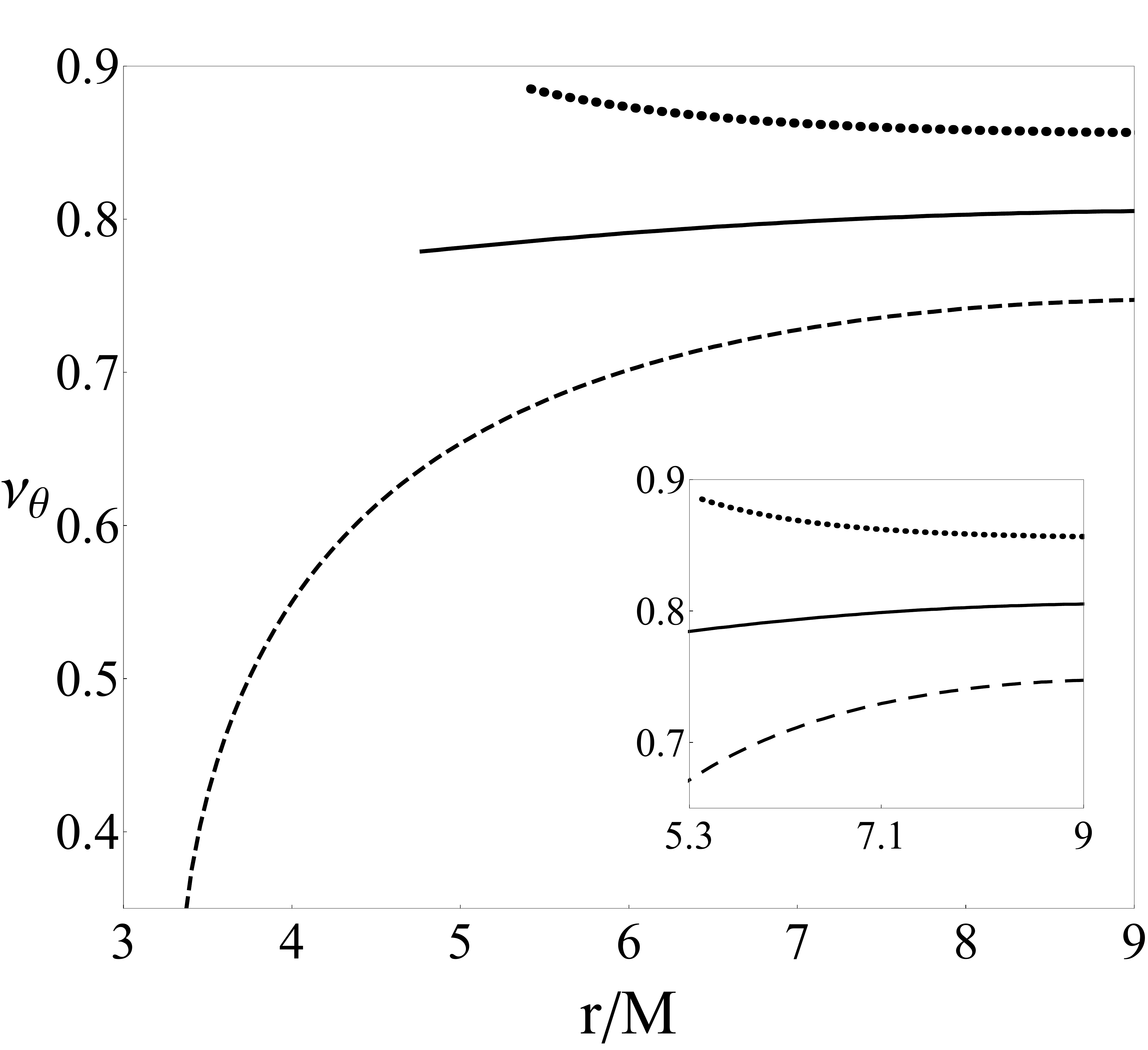}
\caption{The rotation curve as a function of radius along the line $\dot{r}=0$ for a surface of section for bound orbits of a test particle with energy $E=0.95\mu$, angular momentum $L_z=3M\mu$, and mass ratio $\mu/M=10^{-6}$ around a deformed Kerr object with angular momentum $a=0.99M$, $\alpha_Q=0$ and $\alpha_{22}=0$ (solid curve), $\alpha_{22}=-2.5$ (dashed curve) and $\alpha_{22}=2.5$ (dotted curve).}
\label{rotation_a22}
\end{figure}

For $\alpha_Q=0$ though with $\alpha_{22}\neq 0$, the metric \eqref{eq:metric} admits integrable geodesic equations, even though the spacetime geometry is not Kerr. The deformation parameter $\alpha_{22}$ couples to the rotation of the black hole and trivially vanishes in the static limit. The impact of $\alpha_{22}$ on the CZV is similar to that of $\alpha_{Q}$: the CZV volume (for fixed $E$ and $L_{z}$) decreases (increases) when $\alpha_{22}$ is positive (negative). Furthermore, since the Carter symmetry is preserved, the equatorial Poincar\'e map is qualitatively similar to that of a particle orbiting in Kerr spacetime. As expected from the KAM and Poincar\'e-Birkhoff theorems, we find no resonant or chaotic orbits, and no Birkhoff islands of stability form anywhere. The rotation curves presented in Fig. \ref{rotation_a22} demonstrate the absence of plateaus and inflection points. For comparison, we include the rotation curve of geodesics in Kerr spacetime (solid line) in this figure. Aside from the fact that the rotation curve produced by a positive $\alpha_{22}$ (dotted curve) has different monotonicity than the ones of Kerr or those with $\alpha_{22}<0$ (dashed curve), the qualitative features of an integrable system remain. The only distinctive feature of a deformed Kerr black hole relative to Kerr from a dynamical system perspective, therefore, is the modification of the orbital evolution through an adjustment of the metric components $g_{\mu \nu}$ appearing within Hamilton's equations \eqref{constants1}--\eqref{eqtheta}: the only pertinent information one can obtain from rotation curves is that the system is integrable.

\section{Gravitational waves}

In this section, we investigate GW emission from the orbits considered in the previous section. Although the metric \eqref{eq:metric} under consideration is not a solution to any known theory of gravity, we treat the problem in the Einstein-quadrupole approximation to understand the phenomenology of the waveforms. The uncertainties generated at this level of approximation likely eclipse any modified gravity adjustments that might alter the quadrupole formula \cite{will18} (though cf. Refs. \cite{tah19,mas20}). A more sophisticated approach would be to solve the (appropriately generalized, see e.g. \cite{suv19}) Teukolsky equation directly or employ the `kludge' waveforms developed in Ref. \cite{glamp07}. In any case, we emphasize that our goal is not to present realistic waveforms to compare with upcoming LISA data, but rather to investigate whether the imprint of integrability manifests, in principle, within a signal at this level.

Ignoring current-multipole contributions, the radiative component of a metric perturbation at large luminosity distance $R$ from a source $\boldsymbol{T}$ can be written, in the transverse-traceless (TT) gauge, as \cite{thorne80,Canizares}
\begin{equation} \label{eq:metpert}
h^{\TT}_{ij}=\frac{2}{R} \frac {d^2 I_{ij}} {dt^2}
\end{equation}
where $I_{ij}$ is the symmetric and trace-free (STF) mass quadrupole associated to the perturbation,
\begin{equation}
I^{ij}=\left[\int d^3x \,x^i x^j\, T^{tt}(t,x^i)\right]^\text{STF},
\end{equation}
and $t$ is the time measured by the detector. The $tt-$component of the stress-energy tensor for a point-particle with trajectory $\boldsymbol{Z}(t)$ reads \cite{peters69}
\begin{equation}
T^{tt}(t,x^i) = \mu\delta^{(3)} \left[x^i - Z^i(t) \right].
\end{equation}
The Boyer-Lindquist coordinates used to describe the metric \eqref{eq:metric} asymptotically reduce to spherical coordinates, so that suitable Cartesian coordinates
\begin{align} \label{eq:cartesian}
x=r \sin\theta \cos\phi,\,\,\,\,\,\, y=r \sin\theta \sin\phi,\,\,\,\,\,\,z=r \cos\theta,
\end{align} 
can be identified, and related to the position of a space-borne detector at infinity. In reality, of course, the detector is not located at infinity but at some finite $R$, so this prescription is not strictly speaking valid, though has been shown to reasonably approximate  EMRI waveforms produced using more sophisticated approaches \cite{glamp07}.  

\begin{figure*}[t]
\includegraphics[scale=0.19]{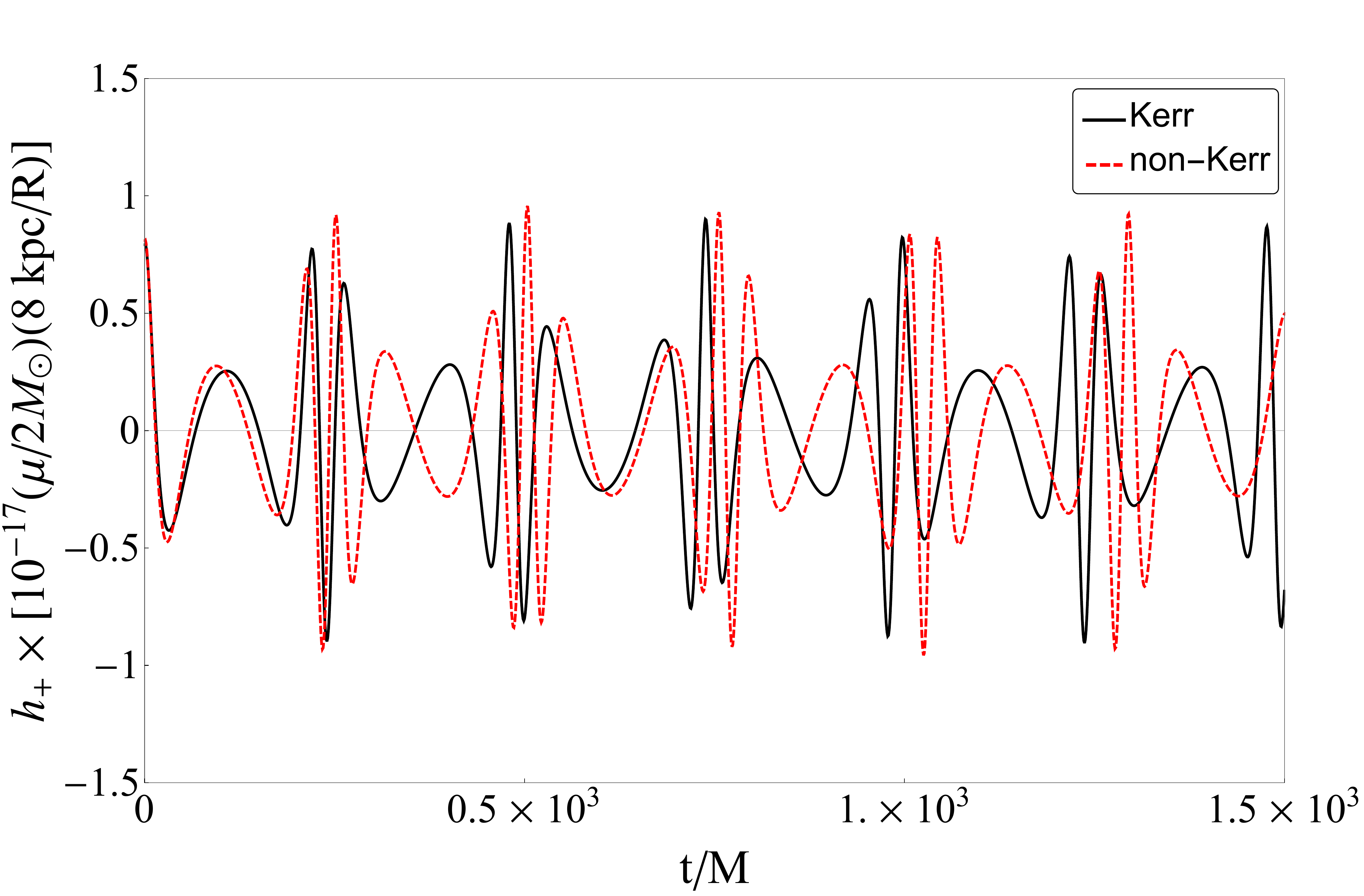}\hskip 1ex
\includegraphics[scale=0.19]{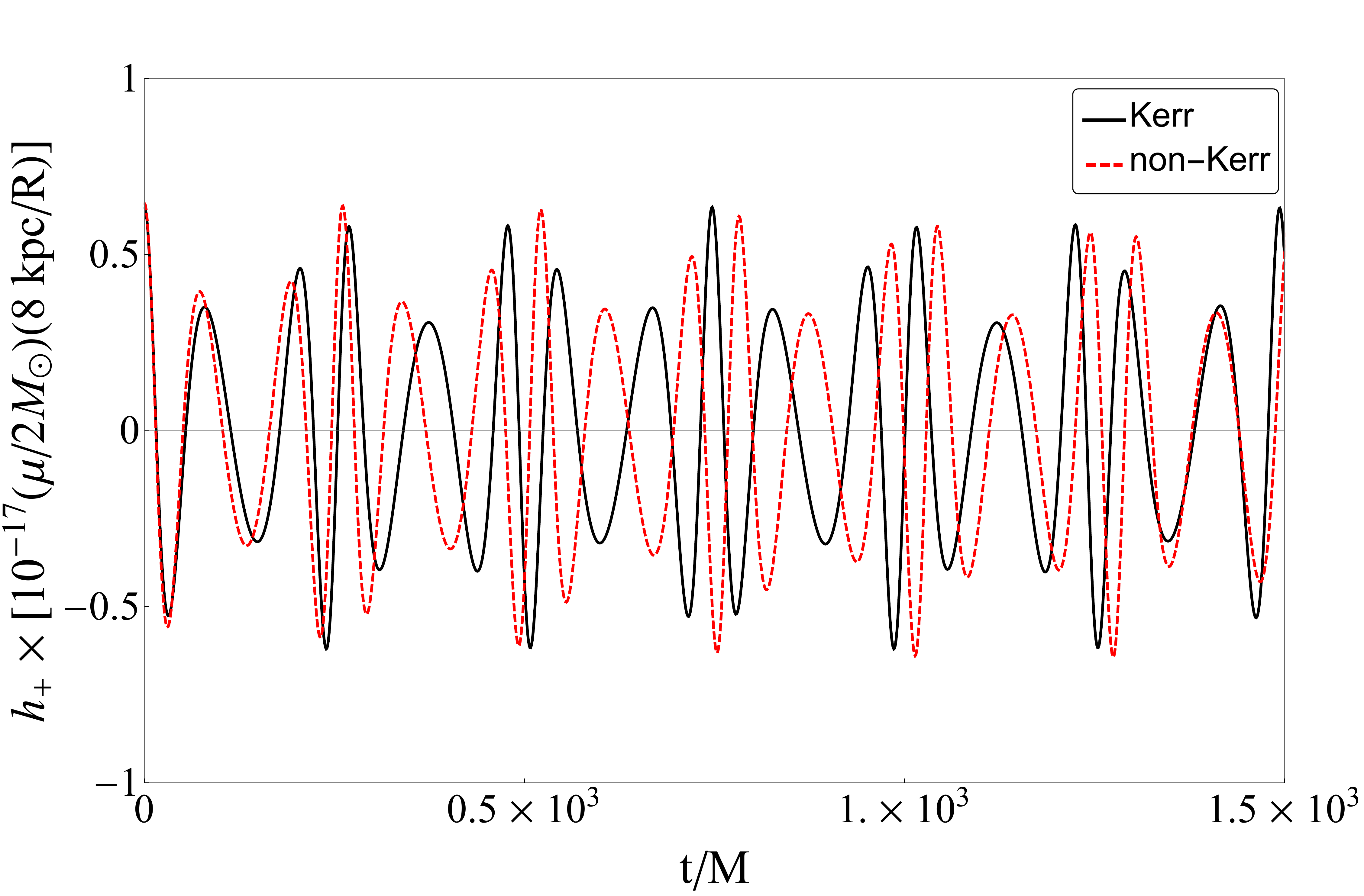}
\includegraphics[scale=0.19]{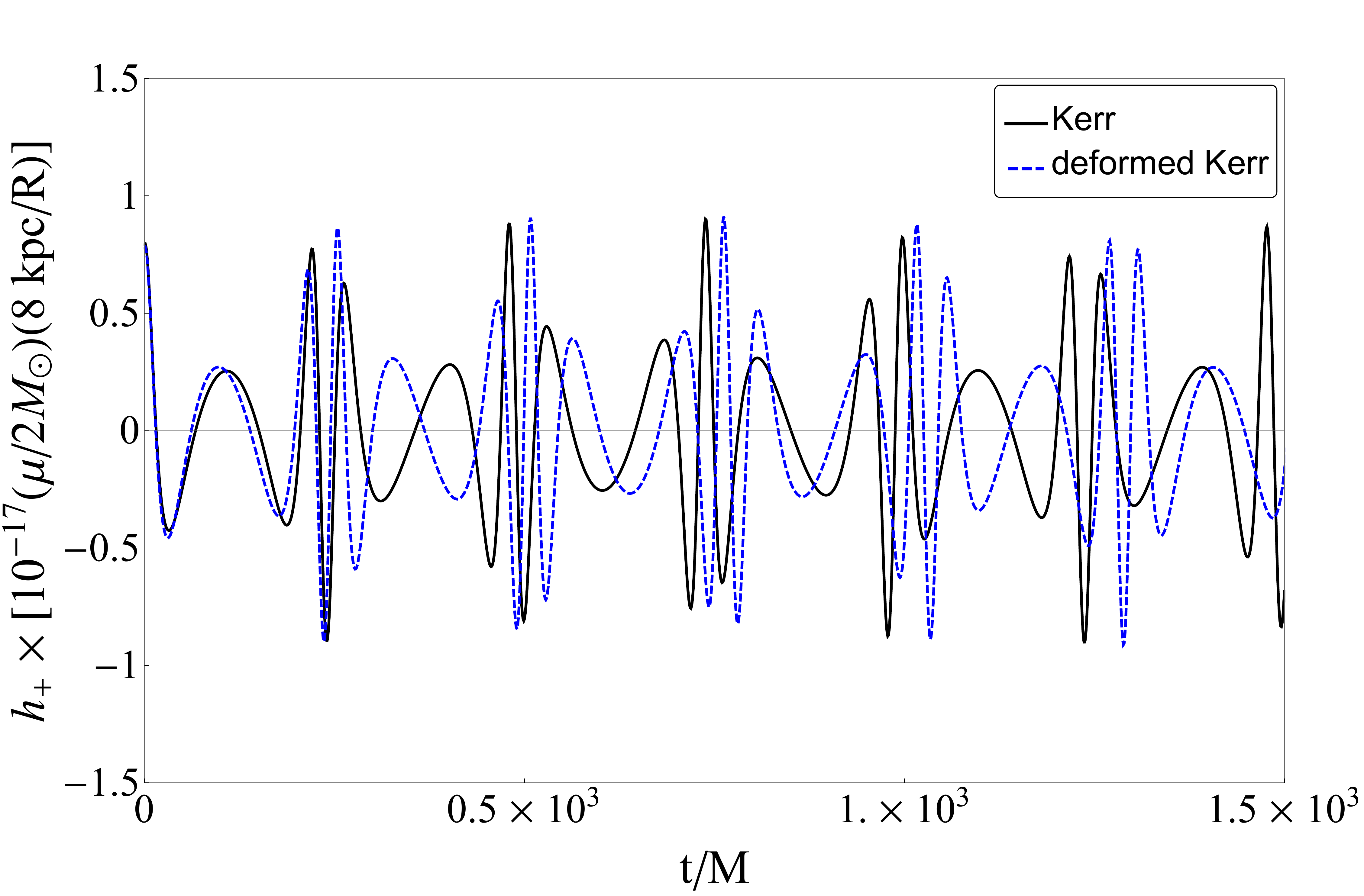}\hskip 1ex
\includegraphics[scale=0.19]{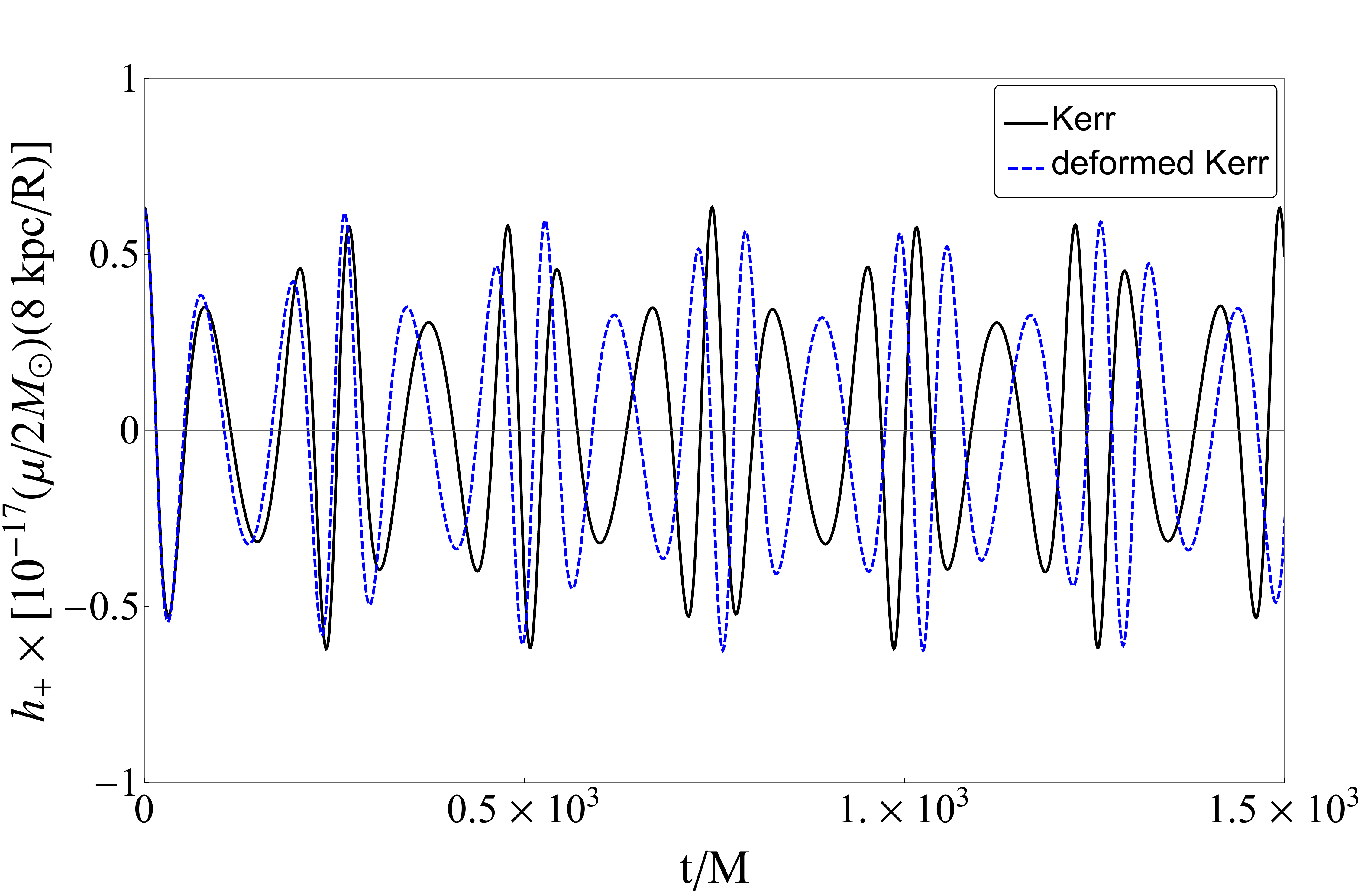}
\caption{Plus polarization waveforms of a test particle with $E=0.95\mu$, $L_z=3M\mu$ and initial conditions $(a)$ (left column) and $(b)$ (right column), orbiting a Kerr black hole (black solid curves), a non-Kerr spacetime with $\alpha_{Q}=-1.8$ (red dashed curves) and a deformed Kerr black hole with $\alpha_{22}=-2.2$ (blue dashed curves) for mass ratio $\mu/M=10^{-6}$ and spin parameter $a=0.99 M$. The source distance is set as $R=8\,\text{kpc}$, while the particle's mass is chosen $\mu=2 M_\odot$ to simulate a stellar mass black hole or a neutron star orbiting around Sgr $\text{A}^{\ast}$.}
\label{GWs}
\end{figure*}

In GR, an incoming GW can be projected onto its mutually orthogonal $+$ and $\times$ polarization states by introducing two vectors $\boldsymbol{p} = \boldsymbol{n} \times {\boldsymbol{Z}} / |\boldsymbol{n} \times {\boldsymbol{Z}}|$ and $\boldsymbol{q} = \boldsymbol{p} \times \boldsymbol{n}$, which are defined in terms of a unit vector $\boldsymbol{n}$ which points from the source to the detector. In terms of the polarization tensors
\begin{equation}
\epsilon_+^{ij}=p^i p^j-q^i q^j,\,\,\,\,\,\,\epsilon_\times^{ij}=p^i q^j+p^j q^i,
\end{equation}
the corresponding GW metric perturbation is
\begin{equation}
h^{ij}(t)=\epsilon_+^{ij}h_+(t)+\epsilon_\times^{ij}h_\times(t),
\end{equation}
with
\begin{equation}
h_+(t)=\frac{1}{2}\epsilon_+^{ij}h_{ij}(t),\,\,\,\,\,\,\,h_\times(t)=\frac{1}{2}\epsilon_\times^{ij}h_{ij}(t).
\end{equation}
In terms of the position, $Z^i(t)$, velocity, $v^i(t)=dZ^i/dt$, and acceleration vectors $a^i(t)=d^2Z^i/dt^2$, one finds \cite{Canizares}
\begin{equation}
\label{GW_formula}
h_{+,\times}(t)=\frac{2\mu}{R}\epsilon^{+,\times}_{ij}\left[a^i(t)Z^j(t)+v^i(t)v^j(t)\right].
\end{equation}

To build some concrete example waveforms, we consider two orbits, with initial conditions 
\begin{enumerate}[$(a)$]
\item: $\left[r(0),\dot{r}(0),\theta(0) \right]=(5.64\,M,\,0.1,\,\pi/2)$,
\item: $\left[r(0),\dot{r}(0),\theta(0) \right]=(7\,M,\,0,\,\pi/2)$,
\end{enumerate}
where we fix the parameters $\mu/M=10^{-6}$, $a=0.99 M$, $E=0.95\mu$, and $L_z=3M\mu$. The remaining initial condition for $\dot{\theta}(0)$ is fixed by the constraint equation \eqref{eq:equationsofmotion}. To test the effect of non-integrability and deformability of the black hole to the gravitational waveform, we take certain choices of the parameters $\alpha_Q$ and $\alpha_{22}$. We remind the reader that a Kerr black hole is recovered for $\alpha_{Q}=\alpha_{22}=0$. We focus on the choice of $\alpha_{Q}=-1.8,\,\alpha_{22}=0$ to simulate bound orbit stages of a non-Kerr EMRI and $\alpha_{22}=-2.2,\,\alpha_Q=0$ to simulate deformed Kerr EMRIs. These choices are made so that the deformed and non-Kerr geodesics, for initial conditions $(a)$ and $(b)$, have the same rotation number (to within $\approx 0.1 \%$). Moreover, we note that the geodesic of case $(a)$ in the non-Kerr spacetime belongs to an island of stability. In both cases, we evolve the geodesic equations for a total time of $t_\text{evol}=2\times 10^6 M\approx 0.3\times \text{year}$, corresponding roughly to $\sim 10^4$ cycles. For simplicity, we fix the location of the detector to be on the $z-$axis at a distance of $R=8 \text{ kpc}$ (the approximate distance to Sgr $\text{A}^{\ast}$) relative to the source. 

In Fig. \ref{GWs} we present the $+$ polarization waveforms for the cases $(a)$ (left panels) and $(b)$ (right panels) discussed above. The top and bottom rows show the non-Kerr (red curve) and deformed-Kerr (blue curve) waveforms, respectively, where we overplot Kerr waveforms (black curves) with vanishing deformation parameters ($\alpha_{Q}=\alpha_{22}=0$) for comparison. For each case, the test particles begin their orbits, on their respective spacetimes, with the same initial conditions. If one compares the waveforms between orbits in the Kerr and non-Kerr spacetimes, their distinction is visually obvious after only half a cycle has passed. The difference between the waveforms initially manifests itself in the form of dephasing (see Sec. V. A below), though after a full cycle develops into something quite marked: while maximum amplitudes are relatively unchanged, peak and trough locations are swapped or significantly shifted relative to Kerr. In any case, a substantial dephasing effect should be evident in upcoming GW data \cite{sop09,Canizares}. A more thorough analysis on the detectability of the parameters $\alpha_Q$ and $\alpha_{22}$ is provided in Sec. V. B below.

\subsection{Frequency spectrum and dephasing}

In this section, we compute frequency-space spectra for the gravitational waveforms shown in Fig. \ref{GWs}. This is achieved by implementing a simple Fourier transform on the underlying data from the numerical geodesic evolution.

\begin{figure*}[t]
\includegraphics[scale=0.24]{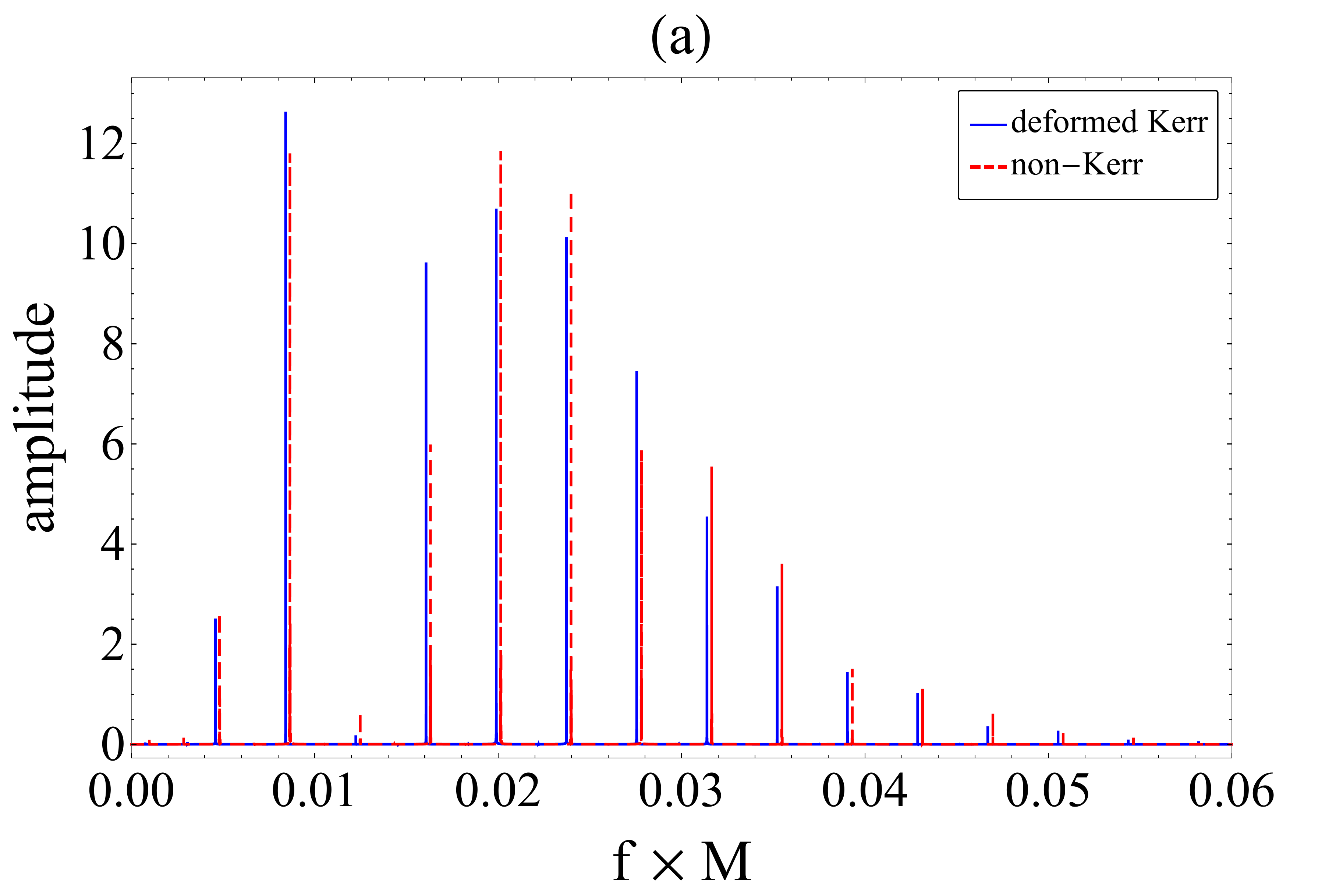}\hskip 1ex
\includegraphics[scale=0.24]{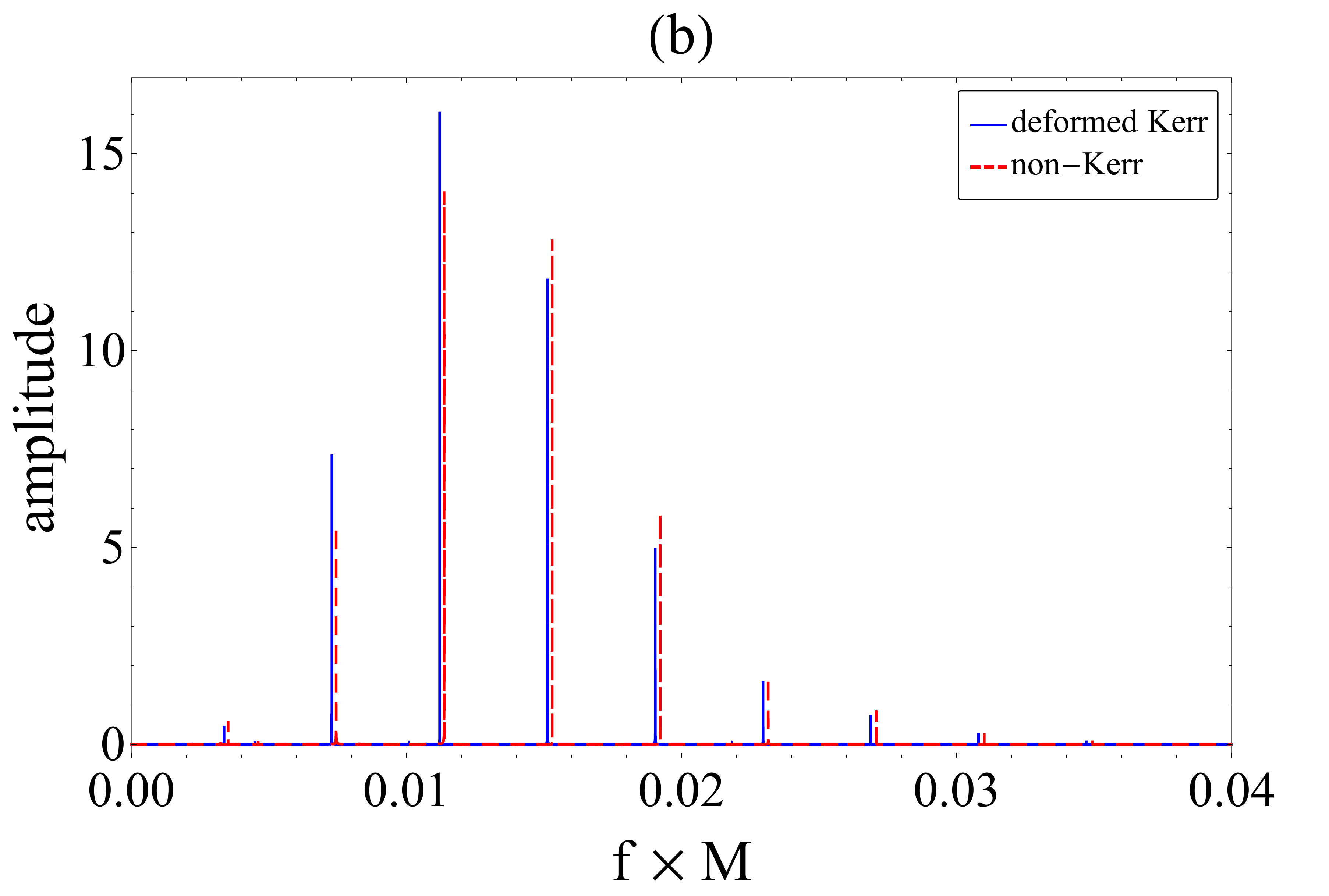}
\includegraphics[scale=0.24]{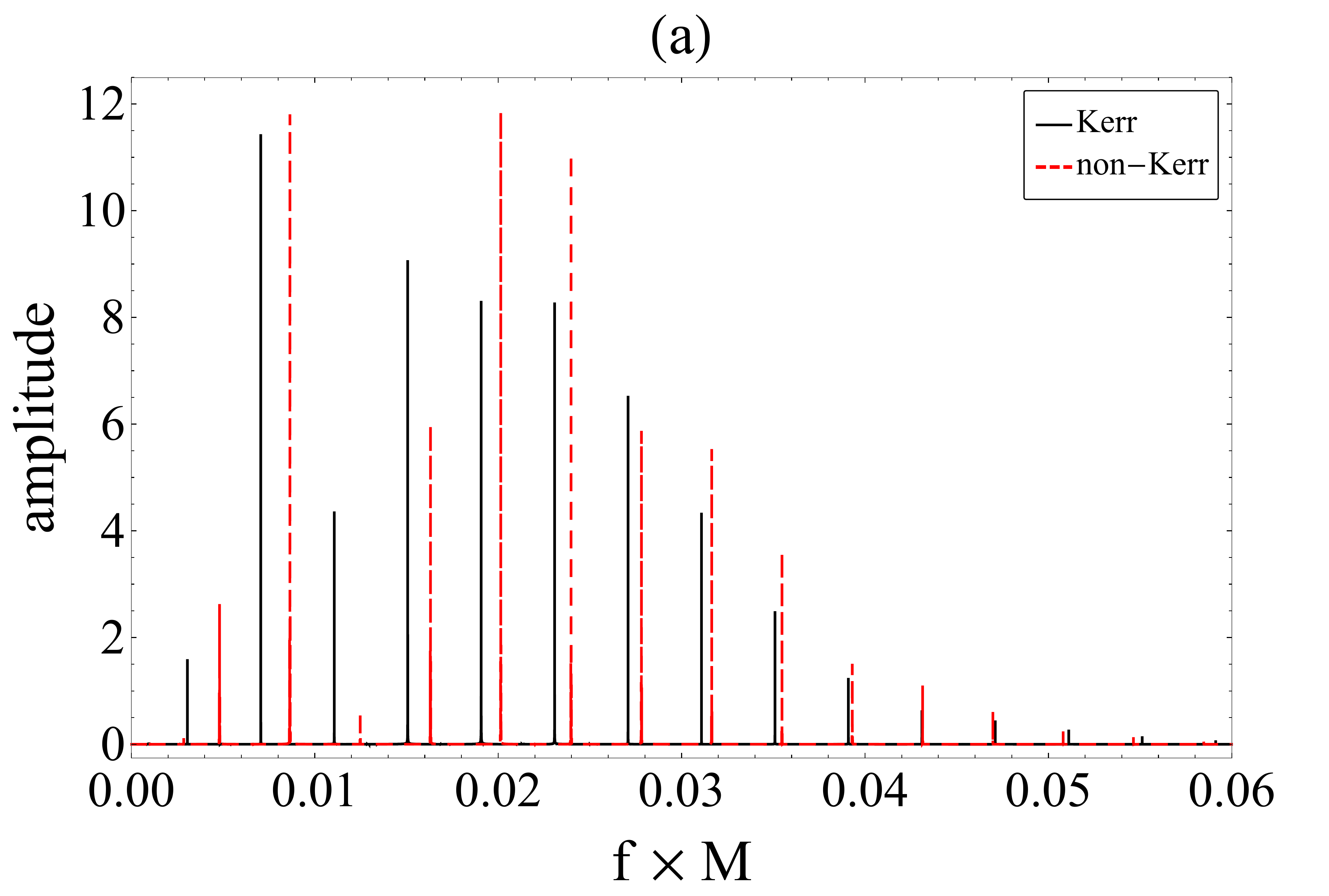}\hskip 1ex
\includegraphics[scale=0.24]{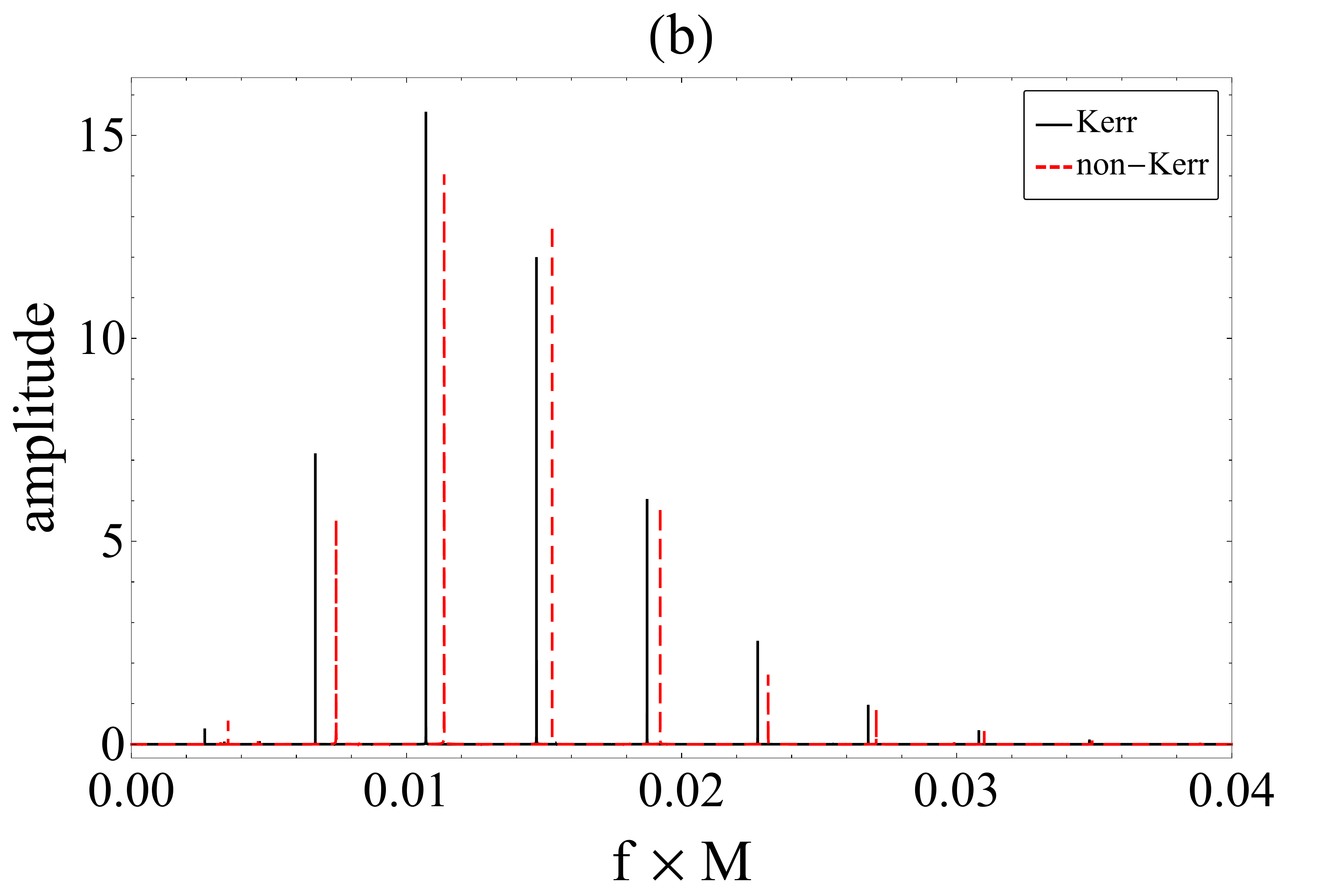}
\caption{Comparison of absolute values of the Fourier transforms of the gravitational wave component $h_+(t)$ 
resulting from Kerr ($\alpha_{22}=\alpha_{Q}=0$), deformed Kerr ($\alpha_{22}=-2.2$) and non-Kerr 
($\alpha_{Q}=-1.8$) EMRI systems, with initial conditions $(a)$ (top and bottom left plots) 
and $(b)$ (top and bottom right plots),
in the frequency domain. For these cases we have fixed the orbital 
parameters as $\mu/M=10^{-6}$, $a=0.99M$, $E=0.95\mu$ and $L_z=3M\mu$. The amplitude 
scaling is arbitrary and the evolution time of the EMRIs is $t_\text{evol}=0.3\times \text{year}$, 
which corresponds roughly to $\sim10^4$ cycles.}
\label{Fourier}
\end{figure*}

Comparative results are shown in Fig. \ref{Fourier}. We see that the frequency shifts relative to Kerr are visually significant, nevertheless the deformed- and non-Kerr cases are remarkably similar (frequency spikes occurring within $\lesssim 0.1 \%$ of each other, which is similar to the relative difference of rotation numbers between those cases). We also observe that the waveforms are multiperiodic (as in \cite{gair08} for geodesics on the Manko-Novikov background), which is an expected feature of generic, non-equatorial orbits \cite{Lewis:2016lgx}.

Despite similarities in the spectra, the frequency shifts seen above lead to a cumulative dephasing in the signals as the orbits evolve. To estimate the extent of dephasing, we employ a Hilbert transform. In general, given a real analytic function $u$, the Hilbert transform, $\mathcal{H}(u)$, is given by
\begin{equation} \label{eq:hilbert}
\mathcal{H}(u)(t) = \frac{1}{\pi} \int^{\infty}_{-\infty} d \tau \frac {u(\tau)} {t - \tau}.
\end{equation}
In particular, the transform is designed so that the complex function $u + i \mathcal{H}(u)$ satisfies the Cauchy-Riemann equations, and is thus holomorphic. Expressing the function in polar form, (e.g. $r_{u} e^{i \theta_{u}}$) therefore, allows for its phase to be read off from the $\theta_{u}(t)$ term within the complex exponential. Given two functions $u_{1}$ and $u_{2}$ which, after taking their respective Hilbert transforms, have polar forms $r_{u_{2}} e^{i \theta_{u_{2}}}$ and $r_{u_{2}} e^{i \theta_{u_{2}}}$, respectively, allows for a determination of the relative phase difference between them: a simple application of de Moivre's theorem gives the phase difference as $\theta_{u_{2}}(t) - \theta_{u_{1}}(t) \equiv \varphi(t)$.

\begin{figure*}[t]
\includegraphics[scale=0.24]{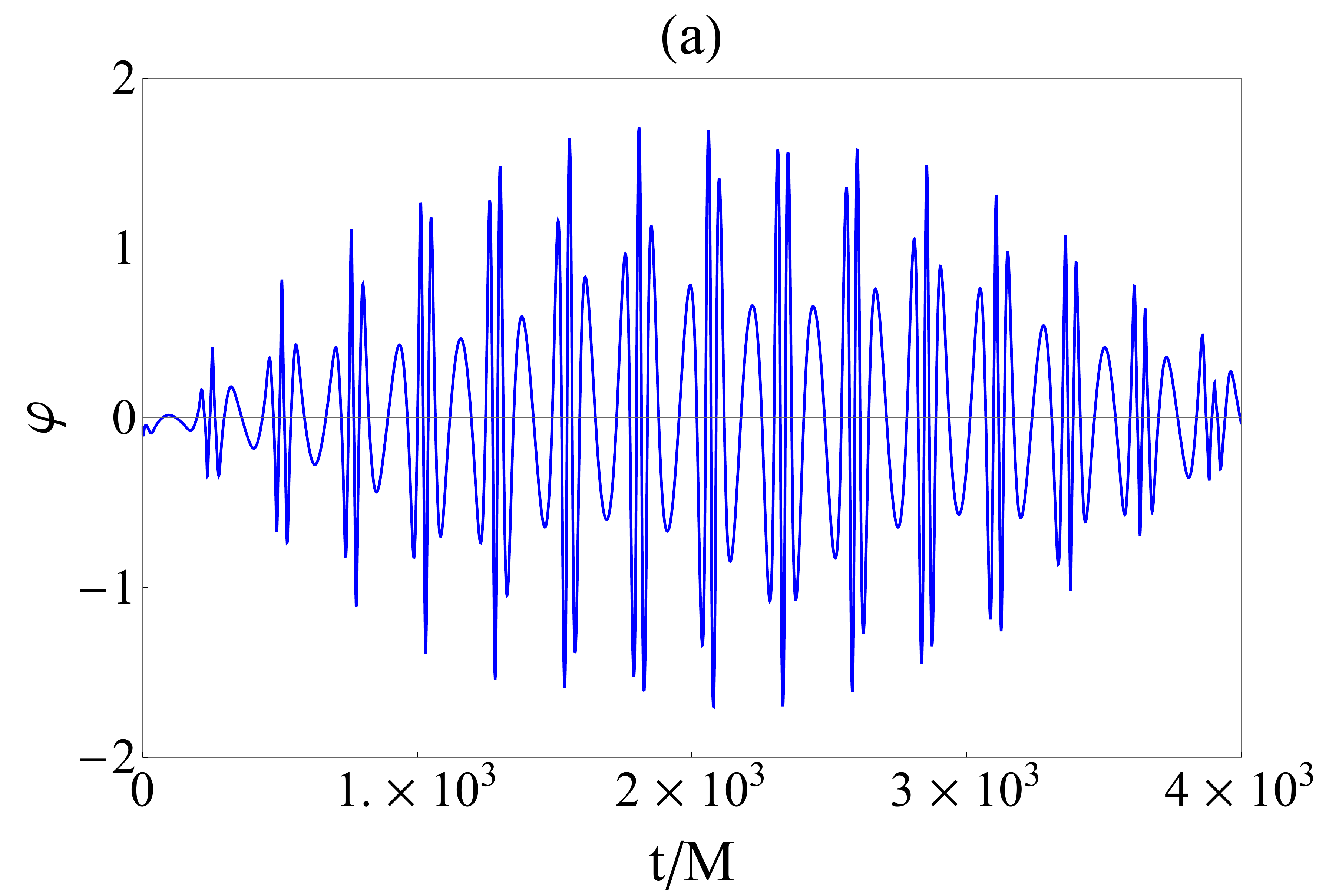}\hskip 1ex
\includegraphics[scale=0.24]{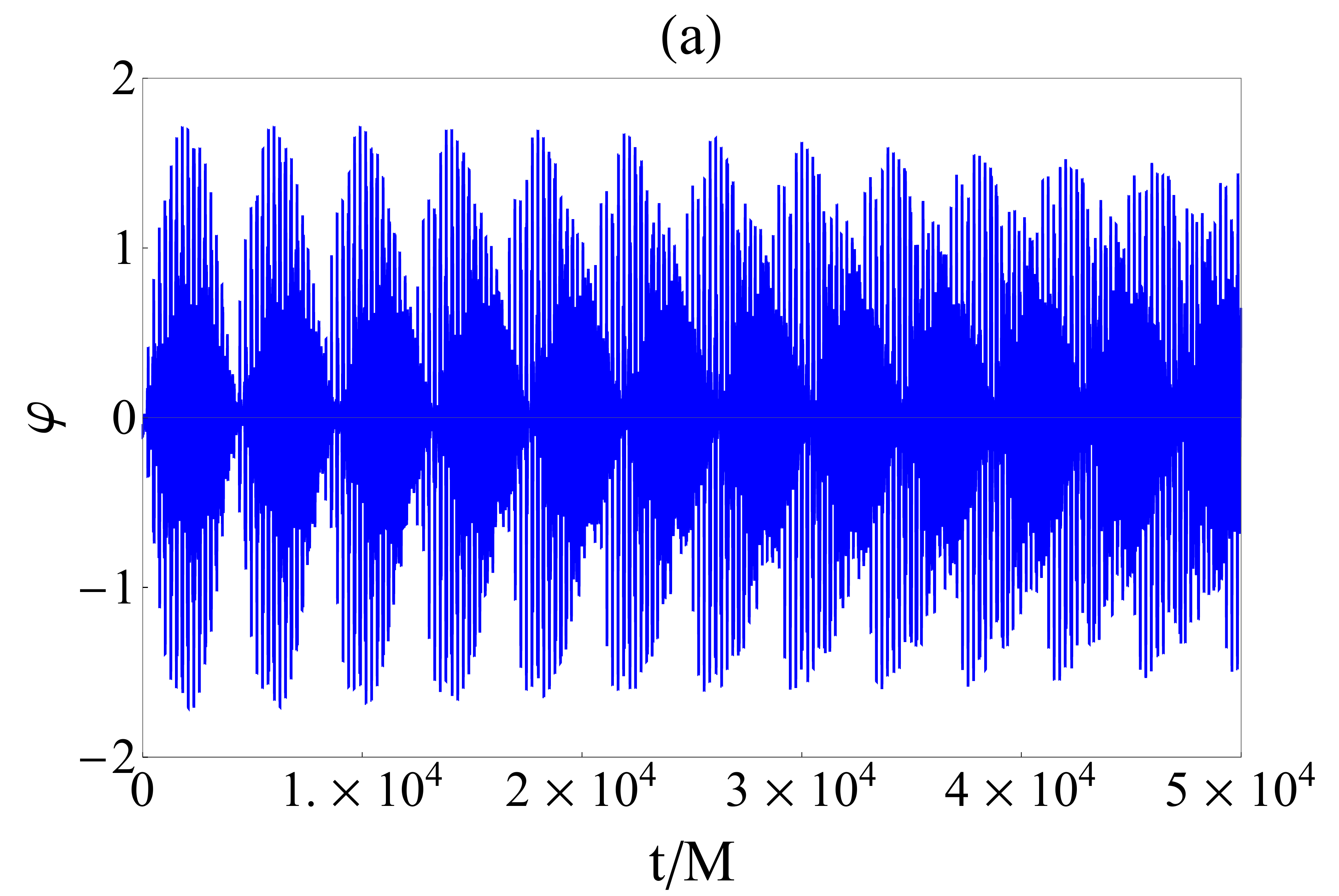}
\includegraphics[scale=0.24]{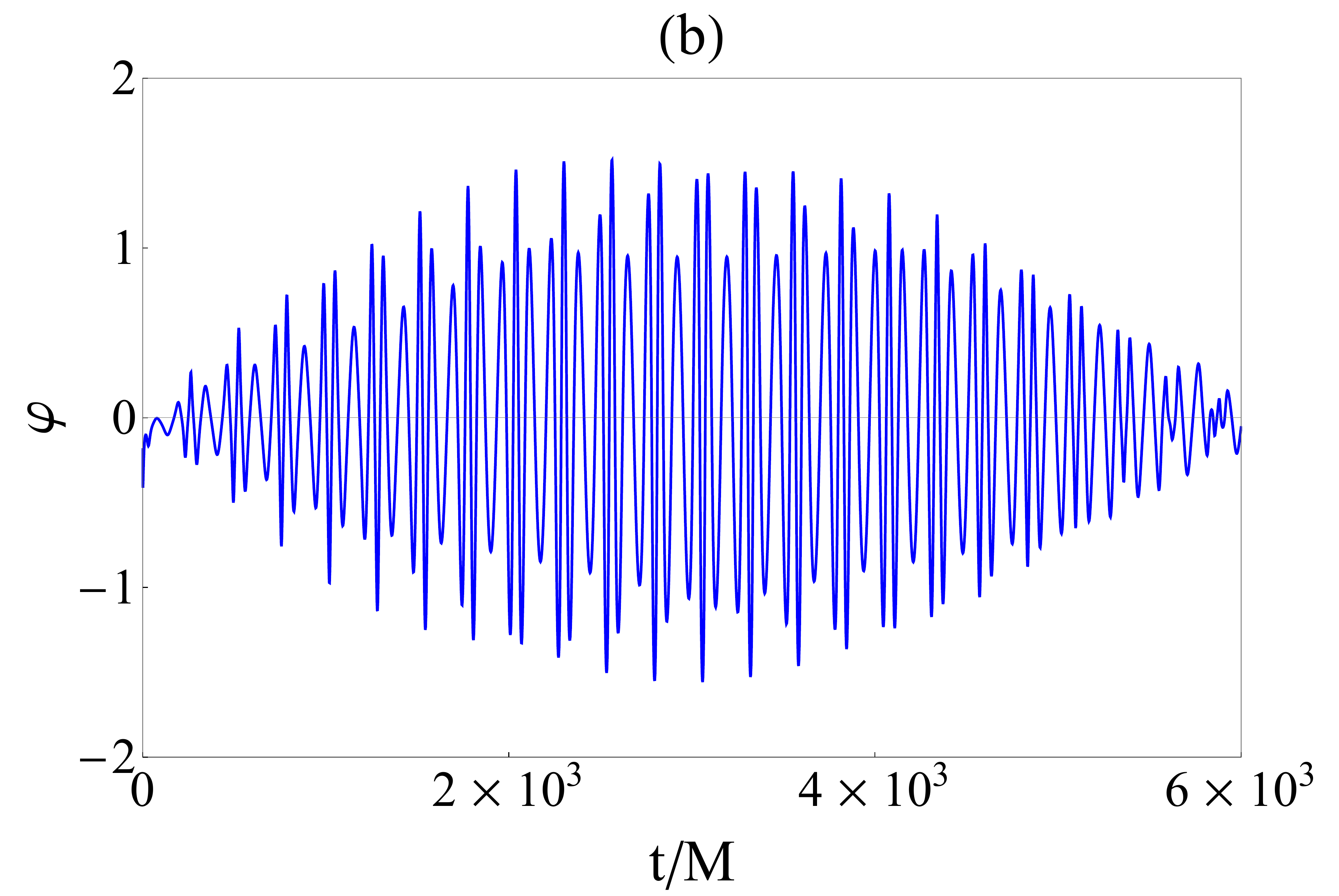}\hskip 1ex
\includegraphics[scale=0.24]{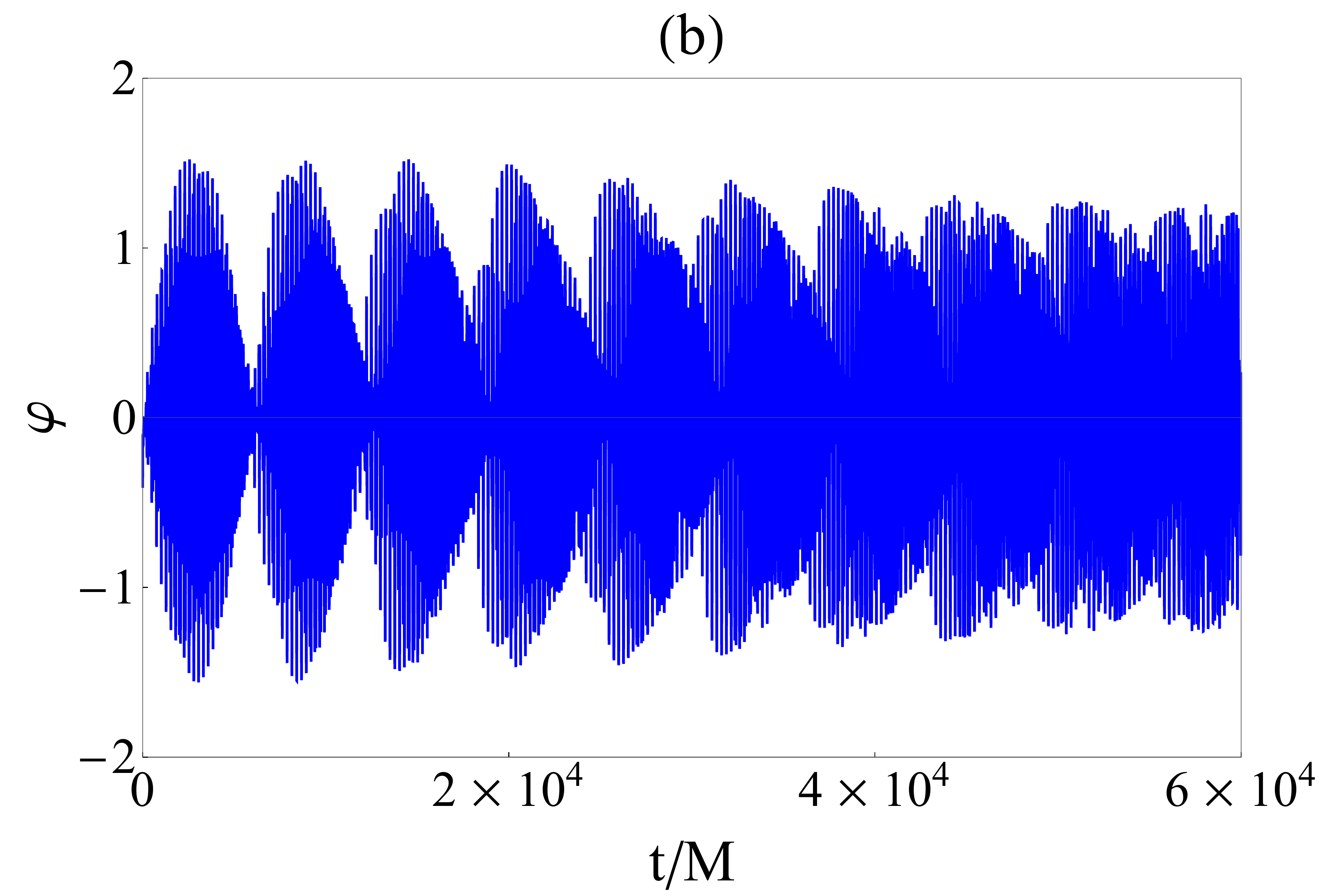}
\caption{Relative phase $\varphi$ between the gravitational wave component $h_+(t)$ of deformed Kerr 
($\alpha_{22}=-2.2$) and non-Kerr ($\alpha_{Q}=-1.8$) EMRI systems with orbital parameters 
$\mu/M=10^{-6}$, $a=0.99M$, $E=0.95\mu$, and $L_z=3M\mu$, and initial conditions $(a)$ (top left and right plots) 
and $(b)$ (bottom left and right plots). The left column displays the relative phases for earlier times while the 
right column displays the relative phases for later times.}
\label{Hilbert}
\end{figure*}

Cumulative dephasings in the above sense are shown in Fig. \ref{Hilbert}. We see that the phase difference between the two signals oscillates rapidly. In both cases $(a)$ and $(b)$, the phase difference is almost periodic for early times (left panels), indicating that the non-Kerr waveform is unlikely to be easily distinguished from the deformed-Kerr counterpart for short observation times. However, we see that for $t \gtrsim 10^{4} M$ (corresponding to $\gtrsim$1 day in real time observation; right panels) the phase difference ceases to be periodic and therefore a significant dephasing accumulates. The rate at which dephasing occurs is sensitive to the initial conditions, as we see that the approximate periodicity in case $(a)$ persists for $\sim$ twice as long. In either case, such a dephasing means that LISA may be able to distinguish between the two spacetimes \cite{sop09}.

\subsection{Parameter estimation}

In this section we provide some estimates for the detectability of the deformed- and non-Kerr parameters by LISA from a hypothetical data stream, $\textbf{s}$. In general, we assume that $\textbf{s}$ includes both some stationary, Gaussian noise, and an EMRI waveform. Provided that the signal-to-noise ratio (SNR) is high enough \cite{vas08}, the prevalance of beyond-Kerr parameters $\lambda^{i}$ can be estimated by computing a relevant Fisher metric built from expectations of random variables associated with $h_{+,\times}$ and the noise within the stream \cite{fisher35,Canizares}. To achieve this, we calculate the pattern response functions relative to the GW polarizations found in Sec. V and use the spectral power densities relevant for LISA. Details of the computation and the assumptions on the noise, LISA's response function, and other relevant aspects are given in Appendix A. 

The critical quantity of interest is ultimately the precision $\Delta \lambda^{i}$ [defined in equation \eqref{eq:fisherest} from the inverse components of the Fisher metric] with which one may measure the parameter vector $\boldsymbol{\lambda} = (\alpha_{Q},\alpha_{22})$ (for a fixed mass and spin). To achieve this in practice, we evolve a large set of orbits for various parameter values and interpolate on them to build `total' waveforms $h_{+,\times}(\alpha_{Q},\alpha_{22},t)$ from which the relevant inner-products and Fourier transforms can be defined.

Table \ref{tab:errorstab} shows relative precisions between the deformed- and non-Kerr parameters for a fixed orbit with initial conditions $(b)$ for various $\boldsymbol{\lambda}$. The relevant ratio of precision $\Delta{\alpha_{Q}}/\Delta{\alpha_{22}}$ is sensitive to the choice of the deformation parameters. When the deformation parameters satisfy $\alpha_{Q}<\alpha_{22}$ then $\alpha_{22}$ can be estimated more accurately than $\alpha_{Q}$, while for $\alpha_{Q}>\alpha_{22}$ we have that $\alpha_{Q}$ can be estimated better than $\alpha_{22}$. Our analysis finds that neither parameter can be estimated better than the other when the deformations are roughly equal $(\alpha_{Q} \approx \alpha_{22})$; the deformation parameters are not `orthogonal' with respect to the Fisher information metric. This latter effect is likely a consequence of the large spin value we have chosen, i.e. $a = 0.99M$. In particular, since $\alpha_{22}$ couples to the spin, its observational prominence scales directly with the value of $a$. Had we considered a slower system, the relative precisions would be skewed in favor of detecting $\alpha_{Q}$, which introduces a deformation to the metric \eqref{eq:metric} even in the static limit. Although supermassive black holes tend to be rapidly rotating, a sizeable number of the highest mass systems with $M \gtrsim 4 \times 10^{7} M_{\odot}$ have intermediate spin values $0.4 \lesssim a \lesssim 0.8$ \cite{rey13}.

\begin{table}
\caption{Ratios of precision with which we are able to measure particular beyond-Kerr parameters. The ratios $\Delta \alpha_{Q} / \Delta \alpha_{22}$ are computed from the inverse entries of the Fisher matrix; see equation \eqref{eq:fisherest}. For these cases we have fixed the orbital parameters as $\mu/M=10^{-6}$, $a=0.99M$, $E=0.95\mu$ and $L_z=3M\mu$ while the initial conditions of the EMRI system are those of case $(b)$. The evolution time of the EMRI is $t_\text{evol}=0.3\times \text{year}$, which corresponds roughly to $\sim10^4$ cycles.}
\begin{tabular}{c|c|c}
\hline
\hline
Parameter $\alpha_{Q}$ & Parameter $\alpha_{22}$ & Precision $ {\Delta \alpha_{Q}} / {\Delta \alpha_{22}}$ \\
\hline
-1.6 & -2 & 1.0671 \\
\hline
-1.8 & -2.2 & 1.1428 \\
\hline
-2 & -1.6 & 0.8165 \\
\hline
-2.2 & -1.8 & 0.7912 \\
\hline
-2 & -2 & 0.9925 \\
\hline
\hline
\end{tabular}
\label{tab:errorstab}
\end{table}



\section{Discussion}

The Kerr metric uniquely describes asymptotically flat and astrophysically stable black holes in GR \cite{Heusler,bhuniq1}. However, several theoretical and observational issues hint at new gravitational physics at the ultraviolet scale \cite{thooft}, implying that astrophysical black holes may have non-Kerr hairs (e.g. \cite{babak05}). These features might theoretically manifest in a number of ways, one interesting possibility of which is that some fundamental symmetries are broken, such as the Carter symmetry which implies the integrability of the geodesic equations \cite{papkok18,car20}. In this paper we explore signatures of spacetime symmetries in orbital dynamics (Sec. III), and investigate whether non-integrability imprints signatures onto GW signals associated with EMRIs for a particular, new metric \eqref{eq:metric} first appearing in this work. Under optimistic astrophysical assumptions, LISA is expected to detect GWs from $\sim 10^{3}$ EMRI events per year \cite{babak17}, and it is therefore important to understand how beyond-Kerr features or spacetime symmetries might evince themselves in upcoming data. 

This work builds on previous studies \cite{apo07,luk10,cont11,luk14,zil20} by introducing a new spacetime metric \eqref{eq:metric}, which contains two free parameters, one of which leads to a symmetry-preserving deformation of the Kerr spacetime, while the other explicitly breaks the Carter symmetry and leads to non-integrable dynamics (see Sec. II). This spacetime admits several desirable properties, such as possessing a horizon and respecting post-Newtonian constraints (see Sec. III), and allows for the ``switching off'' of the Carter symmetry in a simple, precise way. Geodesics within this spacetime were studied numerically, and clear non-integrable signatures in the orbital dynamics were found when the Carter parameter was broken, most notably in the rotation numbers; see Fig. \ref{rotation_aQ}. 

Although the distinction between Kerr and beyond-Kerr geodesics is apparent in our analysis, no obvious imprint is carried over into the gravitational waveforms of deformed and non-Kerr geodesics besides a substantial dephasing (occurring roughly after just a day of real time observation; see Fig. \ref{Hilbert}.). The Fourier spectra (Fig. \ref{Fourier}) of the resulting GWs from deformed and non-Kerr EMRIs, with similar rotation numbers, are practically indistinguishable, and a Fisher metric analysis indicates that both deformation parameters can be approximated with similar precision. This is likely due to our limited prescription. 

Indeed, a full study of a realistic, non-Einstein EMRI problem is beyond the scope of this work, and may be the reason why we only see limited impact. For example, we do not include: radiative backreactions \cite{quinn97,self1,self2}, self gravity or finite-size effects for the companion \cite{finite1,finite2}, current, higher-mass, or post-Newtonian multipole moments (i.e. no Teukolsky equation) \cite{thorne80,moms}, exact Cartesian-like coordinates but rather only asymptotically Cartesian ones when evaluating $\boldsymbol{Z}(t)$ \eqref{eq:cartesian} \cite{glamp07}, Solar system motions or orbital systematics which influence the detector orientation $\boldsymbol{n}$ \cite{Canizares}, non-Einstein corrections for GW formulae \citep{will18,suv19,tah19,mas20}, internal spin or angular momentum for the companion (i.e. no Mathisson-Papapetrou-Dixon equations) \cite{kiu04,spin1,spin2,zil20}, or 3-body interactions with other objects near the super-massive black hole or its accretion disk \cite{rez07,rez08,sem13}. These effects will be incorporated in a future study. In any case, the work presented here demonstrates that beyond testing the validity of GR, GWs from EMRIs may be able to also probe fundamental aspects of spacetime, such as their symmetries.

\section*{Acknowledgements}
We would like to thank George Papadopoulos for discussions. We would also like to thank Georgios Lukes-Gerakopoulos and Theocharis Apostolatos for important feedback and discussions on a previous version of this manuscript.  Thanks are due to the anonymous referee for providing useful feedback, especially in the GW modeling section, which significantly improved the quality of this manuscript. AGS is supported by the Alexander von Humboldt Foundation, and extends his thanks to Prof. Bill Moran for hospitality shown during a research visit to Melbourne, where some of this work was completed.
\appendix

\section{Calculation of the Fisher information metric}

In this appendix, which closely follows the work of Canizares et al. \cite{Canizares} and references therein, we present details of the calculation of the Fisher metric used in Sec. V. B. 

In general, LISA's response to an incident GW is determined by a vector $h_{\alpha}$ which depends on the antenna pattern response functions $F^{+,\times}_{I,II}$ and the plus and cross polarizations through \cite{barack04}
\begin{equation}
h_{\alpha}(t)=\frac{\sqrt{3}}{2}\left[F^{+}_{\alpha}(t)h_{+}(t)+F^{\times}_{\alpha}(t)h_{\times}(t)\right],
\end{equation}
where $\alpha=(I,II)$ is an index representing the different independent channels of the interferometer, and
\begin{eqnarray}
F^{+}_I &=& \frac{1}{2}(1+\cos^2\theta)\cos(2\phi)\cos(2\psi) \nonumber \\
&&-\cos\theta\sin(2\phi)\sin(2\psi), \\
F^{\times}_I &=& \frac{1}{2}(1+\cos^2\theta)\cos(2\phi)\cos(2\psi)\nonumber \\
&&+\cos\theta\sin(2\phi)\sin(2\psi), \\
F^{+}_{II} &=& \frac{1}{2}(1+\cos^2\theta)\sin(2\phi)\cos(2\psi) \nonumber 
\\&&+\cos\theta\cos(2\phi)\sin(2\psi), \\
F^{\times}_{II} &=& \frac{1}{2}(1+\cos^2\theta)\sin(2\phi)\sin(2\psi) \nonumber \\
&&-\cos\theta\cos(2\phi)\cos(2\psi).
\end{eqnarray}

Assuming a fixed orientation $\boldsymbol{n} = (0,0,1)$ and that the primary hole's spin polar and spin azimuthal angles remain fixed at the equatorial plane for simplicity, as in Sec. V, the angles $(\theta,\phi, \psi)$ (not to be confused with Boyer-Lindquist coordinates) introduced above read (see Ref. \cite{Canizares} for formulae in the general case) $\theta(t) = \pi/3$, $\phi(t) = 2 \pi t /T + \pi/2$, and $\psi = - 2 \pi t /T$, where $T=1$ year is the orbital period of the Earth around the Sun.

Suppose that the data stream observed by the detector, $s_{\alpha}(t)$, contains both an EMRI signal $h_{\alpha}(t)$ and some noise $n_{\alpha}(t)$, viz.
\begin{equation}
s_\alpha(t) = h_\alpha(t) + n_\alpha(t).
\end{equation}
To make progress, we assume that $n_{\alpha}$ is both stationary and Gaussian. We further assume that the two data streams (i.e., $\alpha = I$ and $\alpha = II$) are uncorrelated and that the power spectral density of the noise for LISA, $S_{\alpha,n}(f)$, is the same in each channel, $S_{I,n}(f) = S_{II,n}(f)$, so that we may drop the subscript $\alpha$ on this quantity. The Fourier components of the noise are therefore given by
\begin{equation}
\langle \tilde{n}_\alpha (f)\tilde{n}^{\ast}_\beta (f')\rangle = \frac{1}{2} 
\delta^{}_{\alpha\beta} \delta(f-f') S_n(f) ,
\end{equation}
where $\langle \cdot \rangle$ denotes an ``ensemble average'' over all possible realizations, the asterisk denotes complex conjugation, and the Fourier transform is denoted with an overhead tilde. The power spectral density is given explicitly by \cite{barack04}
\begin{equation}
S_n = \text{min}\left\{ S_{n}^{\text{inst}}+  S_{n}^{\text{exgal}}, S_n^{\text{inst}}+ S_{n}^{\text{gal}}+ S_{n}^{\text{exgal}}\right\},
\end{equation}
where $S_{n}^{\text{inst}}(f)$ denotes the instrumental noise while $S_{n}^{\text{gal}}(f)$ and $S_{n}^{\text{exgal}}(f)$ denote `confusion' noises from galactic and extra-galactic binaries, respectively. These quantities have the following functional forms \cite{Canizares}
\begin{equation}
\begin{aligned}
\hspace{-0.4cm}S_{n}^{\text{inst}}(f) =& \exp{\left(\kappa T^{-1}_{\text{mis}} \frac{{\rm d}N}{{\rm d}f}\right)} (9.18\times10^{-52}f^{-4} \\
& + 1.59\times10^{-41} + 9.18\times10^{-38}f^{2}) \text{ Hz}^{-1},
\end{aligned}
\end{equation}
\begin{equation}
S_{n}^{\text{gal}}(f) = 2.1\times 10^{-45} \left(\frac{f}{1\text{Hz}}\right)^{-7/3} \text{ Hz}^{-1},
\end{equation}
and
\begin{equation}
S_{n}^{\text{exgal}}(f) = 4.2\times10^{-47} \left(\frac{f}{1\text{Hz}}\right)^{-7/3} \text{ Hz}^{-1},
\end{equation}
with ${\rm d}N/{\rm d}f \approx 2\times 10^{-3}\left( 1\text{Hz} / f \right)^{11/3}$ representing the number-density of galactic binaries with $\mu \lesssim M_{\odot}$ per unit frequency, $T_{\text{mis}}$ denoting the lifetime of the LISA mission (which we take to be $10$ years), and $\kappa \approx 4.5 $ being the mean number of frequency bins that are lost when each galactic binary is fitted out from the data \cite{barack04,Canizares}.

By virtue of the Gaussian noise assumption, the probability that a given EMRI GW signal $\textbf{h}$ is present within the data stream $\textbf{s}$ reads
\begin{equation} \label{eq:likelihood2}
p(\textbf{s}|\textbf{h}) \sim \exp \left[- \frac {1} {2} (\textbf{s}-\textbf{h}|\textbf{s}-\textbf{h})\right],
\end{equation}
where $(\textbf{a} |\textbf{b})$ denotes the inner product on the vector space of signals associated with the power spectral density $S_{n}(f)$ \cite{cut94},
\begin{equation}
(\textbf{a}|\textbf{b}) = 2\sum_{\alpha} \int_0^\infty df \frac{\tilde{a}^{\ast}_\alpha(f) \tilde{b}_\alpha(f) + \tilde{a}_\alpha(f) \tilde{b}^{\ast}_\alpha(f)}{S_n(f)} . \label{eq:innerproduct}
\end{equation}

In general, the best-fitting waveform $\textbf{h}$ is defined as the one which maximizes $(\textbf{s}|\textbf{h})$. In practice, however, one considers a family of waveform templates that depend on a set of parameters $\lambda^{i}$ (say), and searches for those $\lambda^{i}$ that maximize the probability of a certain noise realization \cite{Canizares}. If the SNR is sufficiently large, the best-fit parameters $\lambda^{i}_{0}$ can safely be assumed to follow a Gaussian distribution that is centered around the true values. Taking $\lambda^{i} = \lambda^{i}_{0} + \delta \lambda^{i}$, we expand expression \eqref{eq:likelihood2} around these best-fit values to find
\begin{equation}
p(\delta \lambda) \sim \exp\left(-\frac{1}{2} \Gamma_{jk} 
\delta\lambda^j\delta\lambda^k\right),
\end{equation}
where
\begin{equation} \label{eq:fisher}
\Gamma_{jk} =\left( \frac{\partial \textbf{h}}{\partial \lambda^j} \Big| \frac{\partial \textbf{h}}{\partial \lambda^k} \right)
\end{equation}
is the Fisher information metric \cite{fisher35}, whose inverse is the covariance matrix for the waveform parameters, viz.
\begin{equation} \label{eq:covariance}
\langle \delta\lambda^j\delta\lambda^k \rangle = \left( \Gamma^{-1}\right)^{jk} \left[1+ \mathcal{O}(\text{SNR}^{-1})\right].
\end{equation}

Finally, using expression \eqref{eq:covariance}, we are in a position to estimate the precision $\Delta \lambda^{i}$ with which one may measure the parameter vector $\lambda^{i}$:
\begin{equation}
\Delta\lambda^{i} \approx \sqrt{ \left( \Gamma^{-1}\right)^{ii} }. \label{eq:fisherest}
\end{equation}

\end{document}